\RequirePackage{etoolbox}
\csdef{input@path}{%
}%

\documentclass[12pt]{article}

\usepackage{amsmath}
\usepackage{amssymb}
\usepackage{amsbsy}
\usepackage{amsthm}
\usepackage{mathtools}
\usepackage{mathbbol}
\usepackage{mathrsfs}
\usepackage{bbm} 
\usepackage{graphicx,psfrag,epsf}
\usepackage{enumerate}
\usepackage{natbib}
\usepackage{url} 
\usepackage{float}
\usepackage[skip=5pt]{caption}
\usepackage{subcaption}
\usepackage[normalem]{ulem}
\usepackage{comment}
\usepackage{hhline}
\usepackage{afterpage}
\usepackage{multirow}
\usepackage{adjustbox}
\usepackage{textcomp}
\usepackage{rotating}
\usepackage{tabularx}
\usepackage{xspace}
\usepackage{xr-hyper}
\usepackage[colorlinks,citecolor=blue,urlcolor=blue,filecolor=blue,backref=page]{hyperref}
\usepackage{color,colortbl,soul}
\usepackage[dvipsnames]{xcolor}
\usepackage{blkarray}
\usepackage{tikz}
\usepackage[boxed, vlined]{algorithm2e}
\usepackage[T1]{fontenc}
\usepackage{arydshln}
\usepackage{scalerel,stackengine}
\usepackage[hmargin=1.0in,vmargin=1.0in]{geometry}

\SetAlCapSkip{.5em}

\definecolor{Gray}{gray}{0.9}

\allowdisplaybreaks

\makeatletter
\renewcommand*\env@matrix[1][*\c@MaxMatrixCols c]{%
	\hskip -\arraycolsep
	\let\@ifnextchar\new@ifnextchar
	\array{#1}}
\makeatother

\RequirePackage{doi}

\newcommand{\iidsim}{\overset{iid}{\sim}} 

\newcommand{\given}{\,|\,}

\stackMath
\newcommand\reallywidehat[1]{%
	\savestack{\tmpbox}{\stretchto{%
			\scaleto{%
				\scalerel*[\widthof{\ensuremath{#1}}]{\kern-.6pt\bigwedge\kern-.6pt}%
				{\rule[-\textheight/2]{1ex}{\textheight}}
			}{\textheight}%
		}{0.5ex}}%
	\stackon[1pt]{#1}{\tmpbox}%
}

\theoremstyle{definition}

\newtheorem{proposition}{Proposition}[section]

\makeatletter
\newcommand*{\addFileDependency}[1]{
  \typeout{(#1)}
  \@addtofilelist{#1}
  \IfFileExists{#1}{}{\typeout{No file #1.}}
}
\makeatother

\def\mathcolor#1#{\@mathcolor{#1}}
\def\@mathcolor#1#2#3{%
	\protect\leavevmode
	\begingroup
	\color#1{#2}#3%
	\endgroup
}

\renewcommand{\tilde}[1]{\widetilde{#1}}

\newcommand{\sigmasq}{\sigma^2}

\newcommand{\bolds}[1]{\boldsymbol{#1}}

\newcommand{\calA}{{\cal A}}
\newcommand{\calD}{{\cal D}}
\newcommand{\calG}{{\cal G}}

\newcommand{\calL}{{\cal L}}

\newcommand{\calP}{{\cal P}}

\newcommand{\calS}{{\cal S}}
\newcommand{\calT}{{\cal T}}
\newcommand{\calU}{{\cal U}}
\newcommand{\calV}{{\cal V}}

\newcommand{\Cov}{\bolds{C}}

\newcommand{\ba}{\bolds{a}}
\newcommand{\bA}{\bolds{A}}
\newcommand{\bb}{\bolds{b}}
\newcommand{\bB}{\bolds{B}}

\newcommand{\bD}{\bolds{D}}
\newcommand{\be}{\bolds{e}}
\newcommand{\bE}{\bolds{E}}

\newcommand{\bF}{\bolds{F}}

\newcommand{\bh}{\bolds{h}}
\newcommand{\bH}{\bolds{H}}
\newcommand{\bbH}{\bolds{\mathcal{H}}}

\newcommand{\bI}{\bolds{I}}

\newcommand{\bK}{\bolds{K}}
\newcommand{\bl}{\bolds{\ell}}
\newcommand{\bL}{\bolds{L}}
\newcommand{\bm}{\bolds{m}}

\newcommand{\bO}{\bolds{O}}

\newcommand{\bR}{\bolds{R}}
\newcommand{\bs}{\bolds{s}}

\newcommand{\bT}{\bolds{T}}
\newcommand{\bu}{\bolds{u}}
\newcommand{\bU}{\bolds{U}}
\newcommand{\bv}{\bolds{v}}
\newcommand{\bV}{\bolds{V}}
\newcommand{\bw}{\bolds{w}}
\newcommand{\bW}{\bolds{W}}
\newcommand{\bx}{\bolds{x}}
\newcommand{\bX}{\bolds{X}}
\newcommand{\by}{\bolds{y}}

\newcommand{\bz}{\bolds{z}}
\newcommand{\bZ}{\bolds{Z}}

\newcommand{\bzero}{\mathbf{0}}

\newcommand{\bbeta}{\bolds{\beta}}
\newcommand{\beps}{\bolds{\varepsilon}}
\newcommand{\btheta}{\bolds{\theta}}
\newcommand{\bSigma}{\bolds{\Sigma}}
\newcommand{\bGamma}{\bolds{\Gamma}}
\newcommand{\bLambda}{\bolds{\Lambda}}
\newcommand{\bDelta}{\bolds{\Delta}}
\newcommand{\bmu}{\bolds{\mu}}
\newcommand{\bnu}{\bolds{\nu}}
\newcommand{\bxi}{\bolds{\xi}}

\newcommand{\bwpa}[1]{\bw_{[{#1}]}}

\newcommand{\pa}[1]{\text{Pa}[{#1}]}

\newcommand{\ch}[1]{\text{Ch}[{#1}]}
\newcommand{\bdiag}{\text{blockdiag}}
\newcommand{\tp}{\widetilde{p}}

\usepackage{algorithmicx}
\usepackage{algpseudocode}

\newcommand{\blind}{0}

\newcommand{\modelname}{\textsc{SpamTree}} 
\newcommand{\modelnames}{\textsc{SpamTrees}}

\newcommand{\repourl}{\if0\blind {\url{github.com/mkln/spamtree}}\fi \if1\blind {\url{github.com/}\texttt{[url redacted in blinded version]}}\fi}

\newcommand{\articletitle}{\bf Spatial Multivariate Trees \\for Big Data Bayesian Regression}

\newcommand{\footremember}[2]{%
    \footnote{#2}
    \newcounter{#1}
    \setcounter{#1}{\value{footnote}}%
}
\newcommand{\footrecall}[1]{%
    \footnotemark[\value{#1}]%
} 

\date{ } 
\begin{document} 

\def\spacingset#1{ 
\renewcommand{\baselinestretch}{#1}\small\normalsize} \spacingset{1}

\if0\blind { 
\title{\articletitle} 
\author{Michele Peruzzi\footremember{alley}{Department of Statistical Science, Duke University} \and David B. Dunson\footrecall{alley}%
} 
\maketitle } \fi

\if1\blind { 
\bigskip 
\bigskip 
\bigskip 
\begin{center}
	{\Large \articletitle} 
\end{center}
\medskip } \fi

\bigskip 
\begin{abstract}
High resolution geospatial data are challenging because standard geostatistical models based on Gaussian processes are known to not scale to large data sizes. While progress has been made towards methods that can be computed more efficiently, considerably less attention has been devoted to big data methods that allow the description of complex relationships between several outcomes recorded at high resolutions by different sensors. Our Bayesian multivariate regression models based on spatial multivariate trees (\modelnames) achieve scalability via conditional independence assumptions on latent random effects following a treed directed acyclic graph. Information-theoretic arguments and considerations on computational efficiency guide the construction of the tree and the related efficient sampling algorithms in imbalanced multivariate settings. In addition to simulated data examples, we illustrate \modelnames\ using a large climate data set which combines satellite data with land-based station data. Source code is available at \repourl.
\end{abstract}

\noindent


\spacingset{1.45}

\section{Introduction} \label{section:introduction} 
It is increasingly common in the natural and social sciences to amass large quantities of geo-referenced data. Researchers seek to use these data to understand phenomena and make predictions via interpretable models that quantify uncertainty taking into account the spatial and temporal dimensions. Gaussian processes (GP) are flexible tools that can be used to characterize spatial and temporal variability and quantify uncertainty, and considerable attention has been devoted to developing GP-based methods that overcome their notoriously poor scalability to large data. The literature on scaling GPs to big data is now extensive. We mention low-rank methods \citep{candelarasmussen05, snelsonghahramani07, gp_predictive_process, frk}; their extensions \citep{lowetal2015, ambikasaran16, huang2018, geoga20}; methods that exploit special structure or simplify the representation of multidimensional inputs---for instance, a Toeplitz structure of the covariance matrix scales GPs to big time series data, and tensor products of scalable univariate kernels can be used for multidimensional inputs \citep{gilboaetal2013, moranwheeler2020, loperetal2020}. These methods may be unavailable or perform poorly in geostatistical settings, which focus on small-dimensional inputs, i.e. the spatial coordinates plus time. In these scenarios,  low-rank methods oversmooth the spatial surface \citep{gp_pp_biasadj}, Toeplitz-like structures are typically absent, and so-called \textit{separable} covariance functions obtained via tensor products poorly characterize spatial and temporal dependence. To overcome these hurdles, one can use covariance tapering and domain partitioning \citep{taper1, taper2, fsa, stein2014, katzfuss_jasa17} or composite likelihood methods and sparse precison matrix approximations \citep{vecchia88, grmfields, block_composite_likelihood}; refer to \cite{sunligenton}, \cite{sudipto_ba17}, \cite{Heaton2019} for reviews of scalable geostatistical methods.

\begin{figure}
    \centering
    \includegraphics[width=.95\textwidth]{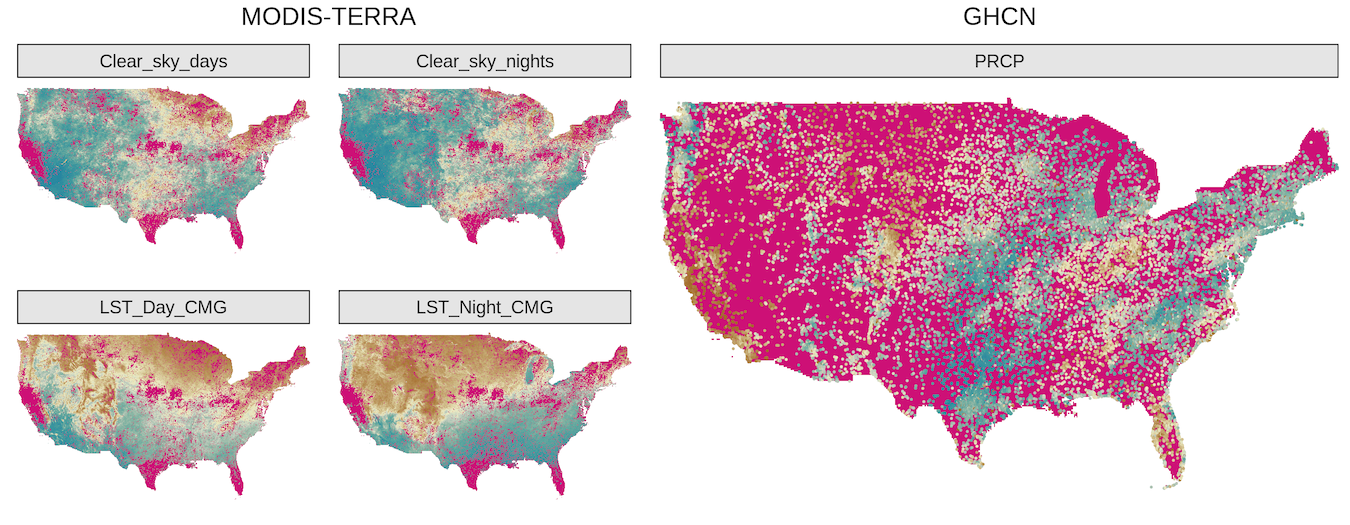}
    \caption{Observed data of Section \ref{section:applications:modisnoaa}. Missing outcomes are in magenta. GHCN data are much more sparsely observed compared to satellite imaging from MODIS.}
\label{fig:data_plot}
\end{figure}

Additional difficulties arise in multivariate (or multi-output) regression settings. Multivariate geostatistical data are commonly misaligned, i.e. observed at non-overlapping spatial locations \citep{spatial_handbook}. Figure \ref{fig:data_plot} shows several variables measured at non-overlapping locations, with one measurement grid considerably sparser than the others. This issue can be solved by modeling cross-dependence of the outputs via latent spatial random effects thought of as a realization of an underlying GP and embedded in a larger hierarchical model.

Unfortunately, GP approximations that do not correspond to a valid stochastic process may inaccurately characterize uncertainty, as the models used for estimation and interpolation may not coincide. Rather than seeking approximations to the full GP, one can develop valid standalone spatial processes by introducing conditional independence across spatial locations as prescribed by a sparse directed acyclic graph (DAG). These models are advantageous because they lead to scalability by construction; in other words, posterior computing algorithms for these methods can be interpreted as approximate algorithms for the full GP, but also as exact algorithms for the standalone process. 

This family of method includes nearest-neighbor Gaussian processes, which limit dependence to a small number of neighboring locations (NNGP; \citealt{nngp, nngp_aoas}), and block-NNGPs \citep{prates}. There is a close relation between DAG structure and computational performance of NNGPs: some orderings may be associated to improved approximations \citep{guinness_techno}, and graph coloring algorithms \citep{molloyreed2002, lewis2016} can be used for parallel Gibbs sampling. Inferring ordering or coloring can be problematic when data are in the millions, but these issues can be circumvented by forcing DAGs with known properties onto the data; in meshed GPs \citep[MGPs;][]{meshedgp}, patterned DAGs associated to domain tiling are associated to more efficient sampling of the latent effects. Alternative so-called multiscale or multiresolution methods correspond to DAGs with hierarchical node structures (trees), which are typically coupled with recursive domain partitioning; in this case, too, efficiencies follow from the properties of the chosen DAG. There is a rich literature on Gaussian processes and recursive partitioning, see e.g  
\cite{ferreira, treed, foxdunson12}; in geospatial contexts, in addition to the GMRF-based method of \cite{nychka15}, multi-resolution approximations \citep[MRA;][]{katzfuss_jasa17} replace an orthogonal basis  decomposition with approximations based on tapering or domain partitioning and also have a DAG interpretation \citep{katzfuss_vecchia}.

Considerably less attention has been devoted to process-based methods that ensure scalability in multivariate contexts, with the goal of modeling the spatial and/or temporal variability of several variables jointly via flexible cross-covariance functions \citep{genton_ccov}. When scalability of GP methods is achieved via reductions in the conditioning sets, including more distant locations is thought to aid in the estimation of unknown covariance parameters \citep{steinetal2004}. However, the size of such sets may need to be reduced excessively when outcomes are not of very small dimension. One could restrict spatial coverage of the conditioning sets, but this works best when data are not misaligned, in which case all conditioning sets will include outcomes from all margins; this cannot be achieved for misaligned data, leading to pathological behavior. Alternatively, one can model the multivariate outcomes themselves as a DAG; however this may only work on a case-by-case basis. Similarly, recursive domain partitioning strategies work best for data that are measured uniformly in space as this guarantees similarly sized conditioning sets; on the contrary, recursive partitioning struggles in predicting the outcomes at large unobserved areas as they tend to be associated to the small conditioning sets making up the coarser scales or resolutions.

In this article, we solve these issues by introducing a Bayesian regression model that encodes spatial dependence as a latent spatial multivariate tree (\modelname); conditional independence relations at the \textit{reference} locations are governed by the branches in a treed DAG, whereas a map is used to assign all \textit{non-reference} locations to leaf nodes of the same DAG. This assignment map controls the nature and the size of the conditioning sets at all locations; when severe restrictions on the reference set of locations become necessary due to data size, this map is used to improve estimation and predictions and overcome common issues in standard nearest-neighbor and recursive partition methods while maintaining the desirable recursive properties of treed DAGs.
Unlike methods based on defining conditioning sets based solely on spatial proximity, \modelnames\ scale to large data sets without excessive reduction of the conditioning sets. Furthermore, \modelnames\ are less restrictive than methods based on recursive partitioning and can be built to guarantee similarly-sized conditioning sets at all locations.

\begin{figure}
	\centering 
		\includegraphics[width=.32\textwidth]{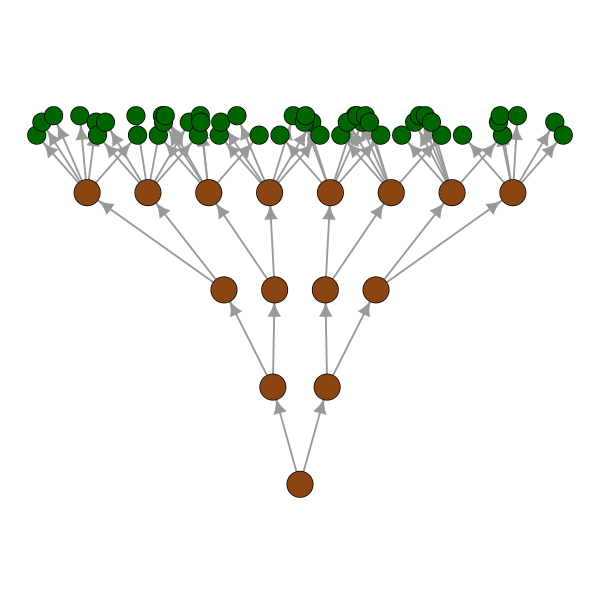}
		\includegraphics[width=.32\textwidth]{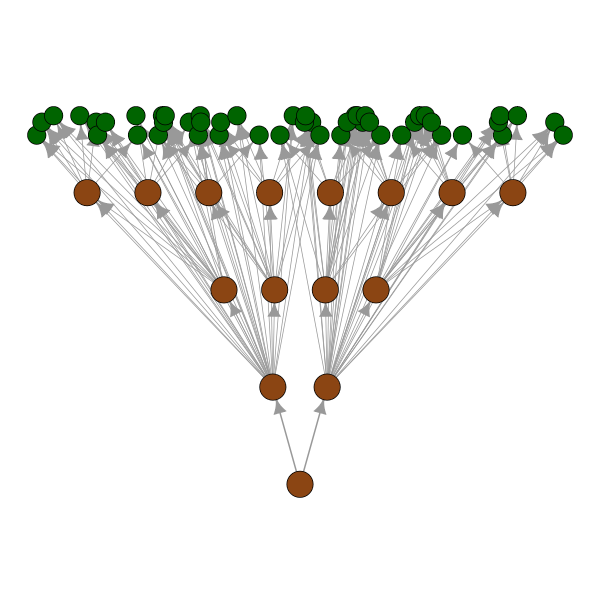} 
		\includegraphics[width=.32\textwidth]{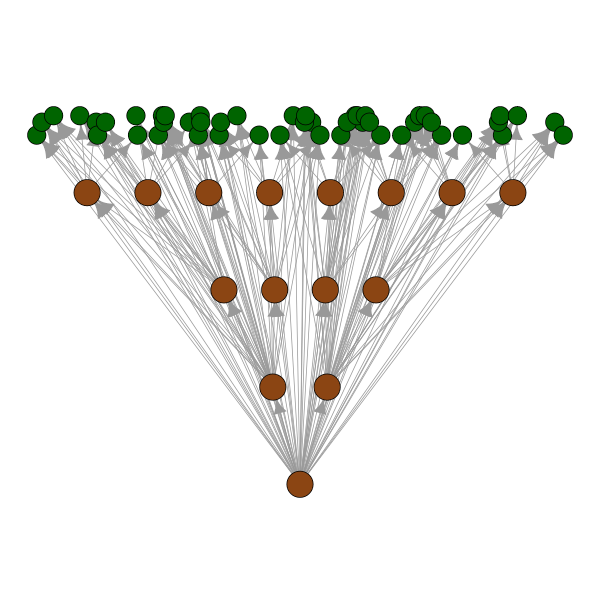} 
		\caption{Three \modelnames\ on $M=4$ levels with depths $\delta=1$ (left), $\delta=3$ (center), and $\delta=4$ (right). Nodes are represented by circles, with branches colored in brown and leaves in green. } 
	\label{figure:spamtree} 
\end{figure}

The present work adds to the growing literature on spatial processes defined on DAGs by developing a method that targets efficient computations of Bayesian multivariate spatial regression models. \modelnames\ share similarities with MRAs \citep{katzfuss_jasa17}; however, while MRAs are defined as a basis function expansion, they can be represented by a treed graph of a \modelname\ with full ``depth'' as defined later (the DAG on the right of Figure \ref{figure:spamtree}), in univariate settings, and ``response'' models. All these restrictions are relaxed in this article. In considering spatial proximity to add ``leaves'' to our treed graph, our methodology also borrows from nearest-neighbor methods \citep{nngp}. However, while we use spatial neighbors to populate the conditioning sets for non-reference locations, the same cannot be said about reference locations for which the treed graph is used instead. Our construction of the \modelname\ process also borrows from MGPs on tessellated domains \citep{meshedgp}; however, the treed DAG we consider here induces markedly different properties on the resulting spatial process owing to its recursive nature. Finally, a contribution of this article is in developing self-contained sampling algorithms which, based on the graphical model representation of the model, will not require any external libraries.

The article builds \modelnames\ as a standalone process based on a DAG representation in Section \ref{section:construction}. A Gaussian base process is considered in Section \ref{section:spamtree_gp} and the resulting properties outlined, along with sampling algorithms. Simulated data and real-world applications are in Section \ref{section:applications}; we conclude with a discussion in Section \ref{section:discussion}. The Appendix provides more in-depth treatment of several topics and additional algorithms.

\section{Spatial Multivariate Trees}\label{section:construction} 
Consider a spatial or spatiotemporal domain $\calD$. With the temporal dimension, we have $\calD \subset \Re^{d} \times [0, \infty)$, otherwise $\calD \subset \Re^{d}$. A $q$-variate spatial process is defined as an uncountable set of random variables $\{\bw(\bl) : \bl\in\calD\}$, where $\bw(\bl)$ is a $q\times 1$ random vector with elements $w_i(\bl)$ for $i=1,2,\ldots,q$, paired with a probability law $P$ defining the joint distribution of any finite sample from that set. Let $\{\bl_1,\bl_2, \ldots, \bl_{n_{\calL}}\} = \calL \subset \calD$ be of size $n_{\calL}$. The $n_{\calL}q\times 1$ random vector $\bw_{\calL} = (\bw(\bl_1)^{\top},\bw(\bl_2)^{\top},\ldots,\bw(\bl_{n_{\calL}})^{\top})^{\top}$ has joint density $p(\bw_{\calL})$. After choosing an arbitrary order of the locations, $p(\bw_{\calL}) = \prod_{i=1}^{n_{\calL}} p( \bw(\bl_{i}) \given \bw(\bl_1), \ldots, \bw(\bl_{i-1})),$ where the conditioning set for each $\bw(\bl_i)$ can be interpreted as the set of nodes that have a directed edge towards $\bw(\bl_i)$ in a DAG. Some scalable spatial processes result from reductions in size of the conditioning sets, following one of several proposed strategies \citep{vecchia88, steinetal2004, gramacy_apley14, nngp, katzfuss_vecchia, meshedgp}. Accordingly, 
\begin{align}\label{eq:dag_basic_multivar} 
	p(\bw_{\calL}) &= \prod_{i=1}^{n_{\calL}} p( \bw(\bl_{i}) \given \bw(\pa{\bl_i})), 
\end{align}
where $\pa{\bl_i}$ is the set of spatial locations that correspond to directed edges pointing to $\bl_i$ in the DAG. If $\pa{\bl_i}$ is of size $J$ or less for all $i=1, \dots, n_{\calL}$, then $\bw(\pa{\bl_i}))$ is of size $Jq$. Methods that rely on reducing the size of parent sets are thus negatively impacted by the dimension $q$ of the multivariate outcome; if $q$ is not very small, reducing the number of parent locations $J$ may be insufficient for scalable computations. As an example, an NNGP model has $\pa{\bl_i} = N(\bl_i)$, where $N(\cdot)$ maps a location in the spatial domain to its neighbor set. It is customary in practice to consider $Jq = m \leq 20$ for accurate and scalable estimation and predictions in univariate settings, but this may be restrictive in some multivariate settings as one must reduce $J$ to maintain similar computing times, possibly harming estimation and prediction accuracy. 

We represent the $i$th component of the $q\times 1$ vector $\bw(\bl)$ as $w(\bl, \xi_i)$, where $\xi_i = (\xi_{i1}, \dots, \xi_{ik})^\top \in \Xi$ for some $k$ and $\Xi$ serves as the $k$-dimensional latent spatial domain of variables. The $q$-variate process $\bw(\bl)$ is thus recast as $\{ w(\bl, \xi) : (\bl, \xi) \in \calD \times \Xi \}$, with $\xi$ representing the latent location in the domain of variables. We can then write (\ref{eq:dag_basic_multivar}) as 
\begin{align}\label{eq:dag_basic_latentrep} 
	p(\bw_{\calL^*}) &= \prod_{i=1}^{n_{\calL^*}} p( w(\bl^*_{i}) \given w(\pa{\bl^*_i})), 
\end{align}
where $\calL^* = \{ \bl^*_i \}_{i=1}^{n_{\calL^*}}$, $\bl^*_i \in \calD \times \Xi = \calD^*$, and $w(\cdot)$ is a univariate process on the expanded domain $\calD^*$. This representation is useful as it provides a clearer accounting of the assumed conditional independence structure of the process in a multivariate context.

\subsection{Constructing spatial multivariate DAGs} \label{section:construction_parts}
\modelnames\ are defined by \textit{(i)} a treed DAG $\calG$ with \textit{branches} and \textit{leaves} on $M$ levels and with depth $\delta \leq M$; \textit{(ii)} a reference set of locations $\calS$; \textit{(iii)} a \textit{cherry picking} map. The graph is $\calG = \{ \bV, \bE \}$ where the nodes are $\bV= \{\bv_1, \dots, \bv_{m_V}\} = \bA \cup \bB$, $\bA \cap \bB = \emptyset$. The \textit{reference} or \textit{branch} nodes are $\bA = \{ \ba_1, \dots, \ba_{m_A} \} = \bA_0 \cup \bA_1 \cup \dots \cup \bA_{M-1}$, where $\bA_i = \{ \ba_{i, 1}, \dots, \ba_{i, m_i} \}$ for all $i=0, \dots, M-1$ and with $\bA_i \cap \bA_j = \emptyset$ if $i\neq j$. The \textit{non-reference} or \textit{leaf} nodes are $\bB = \{ \bb_1, \dots, \bb_{m_B} \}$, $\bA \cap \bB = \emptyset$. We also denote $\bV_r = \bA_r$ for $r=0, \dots, M-1$ and $\bV_{M} = \bB$. The edges are $\bE = \{ \pa{\bv} \subset \bV : \bv \in \bV \}$ and similarly $\ch{\bv} = \{ \bv' \in \bV : \bv \in \pa{\bv'} \}$. The reference set $\calS$ is partitioned in $M$ levels starting from zero, and each level is itself partitioned into reference subsets: $\calS=\cup_{r=0}^{M-1} \calS_r = \cup_{r=0}^{M-1} \cup_{i=1}^{m_i} S_{ri}$, where $S_{ri} \cap S_{r'i'} = \emptyset$ if $r\neq r'$ or $i \neq i'$ and its complement set of \textit{non-reference} or \textit{other} locations $\calU = \calD^* \setminus \calS$. The \textit{cherry picking} map is $\eta: \calD^* \rightarrow \bV$ and assigns a node (and therefore all the edges directed to it in $\calG$) to any location in the domain, following a user-specified criterion. 

\subsubsection{Branches and leaves}
For a given $M$ and a depth $\delta \leq M$, we impose a treed structure on $\calG$ by assuming that if $\bv \in \bA_i$ and $i>M-\delta = M_{\delta}$ then there exists a sequence of nodes $\{\bv_{r_{M_{\delta}}}, \dots, \bv_{r_{i-1}}\}$ such that $\bv_{r_{j}} \in \bA_{j}$ for $j = M_{\delta}, \dots, i-1$ and $\pa{\bv} = \{\bv_{r_{M_{\delta}}}, \bv_{r_1}, \dots, \bv_{r_{j-1}} \}$. If $i\leq M-\delta = M_{\delta}$ then $\pa{\bv} = \{ \bv_{i-1} \}$ with $\bv_{i-1} \in \bA_{i-1}$.
$\bA_0$ is the tree \textit{root} and is such that $\pa{\bv_0} = \emptyset$ for all $\bv_0 \in \bA_0$. The depth $\delta$ determines the number of levels of $\calG$ (from the top) across which the parent sets are nested. Choosing $\delta=1$ implies that all nodes have a single parent; choosing $\delta=M$ implies fully nested parent sets (i.e. if $\bv_i \in \pa{\bv_j}$ then $\pa{\bv_i} \subset \pa{\bv_j}$ for all $\bv_i, \bv_j \in \bV$). The $m_i$ elements of $\bA_i$ are the branches at level $i$ of $\calG$ and they have $i-M_{\delta}$ parents if the current level $i$ is above the depth level $M_{\delta}$ and 1 parent otherwise. We refer to \textit{terminal branches} as nodes $\bv \in \bA$ such that $\ch{\bv} \subset \bB$. For all choices of $\delta$, $\bv \in \bA_i, \bv'\in \bA_j$ and $\bv \in \pa{\bv'}$ implies $i<j$; this guarantees acyclicity.

As for the leaves, for all $\bv \in \bB$ we assume $\pa{\bv} = \{\bv_{r_{M_{\delta}}}, \dots, \bv_{r_{k}}\}$ for some integer sequence $\{ r_{M_{\delta}}, \dots, r_{k} \}$ and $\bv_{r_i} \in \bA_i$ with $i\geq M_{\delta}$. We allow the existence of multiple leaves with the same parent set, i.e. there can be $k$ and $\bb_{i_1}, \dots, \bb_{i_k}$ such that for all $i_2, \dots, i_k$, $\pa{\bb_{i_h}} = \pa{\bb_{i_1}}$. Acyclicity of $\calG$ is maintained as leaves are assumed to have no children. Figure \ref{figure:spamtree} represents the graph associated to \modelnames\ with different depths.

\subsubsection{Cherry picking via $\eta(\cdot)$}
The link between $\calG$, $\calS$ and $\calU$ is established via the map $\eta : \calD^* \rightarrow \bV$ which associates a node in $\calG$ to any location $\bl^*$ in the expanded domain $\calD^*$: 
\begin{align}\label{eq: map} 
	\eta(\bl^*) &= \left\{ 
	\begin{array}{l}
		\eta_{A}(\bl^*) = \ba_{ri} \in \bA_r\; \mbox{ if }\; \bl^* \in S_{ri},\\
		\eta_{B}(\bl^*) = \bb \in \bB \; \mbox{ if }\; \bl^* \in \calU. 
	\end{array}
	\right. 
\end{align}
This is a many-to-one map; note however that all locations in $S_{ij}$ are mapped to $\ba_{ij}$: by calling $\eta(X) = \{ \eta(\bl^*) : \bl^* \in X \}$ then for any $i=0, \dots, M-1$ and any $j=1, \dots, m_i$ we have $\eta(S_{ij}) = \eta_A(S_{ij}) = \ba_{ij}$. \modelnames\ introduce flexibility by cherry picking the leaves, i.e. using $\eta_B : \calU \rightarrow \bB$, the restriction of $\eta$ to $\calU$. Since each leaf node $\bb_j$ determines a unique path in $\calG$ ending in $\bb_j$, we use $\eta_B$ to assign a convenient parent set to $w(\bu)$, $\bu \in \calU$, following some criterion. 

For example, suppose that $\bu = (\bl, \xi_s)$ meaning that $w(\bu) = w(\bl, \xi_s)$ is the realization of the $s$-th variable at the spatial location $\bl$, and we wish to ensure that $\pa{w(\bu)}$ includes realizations of the same variable.  Denote $\bT = \{\bv \in \bA : \ch{\bv} \subset \bB \}$ as the set of terminal branches of $\calG$. Then we find $(\bl, \xi_s)_{\text{opt}} = \arg\min_{(\bl',\xi'=\xi_s) \in \eta_A^{-1}(\bT)} d( \bl' , \bl )$ where $d(\cdot, \cdot)$ is the Euclidean distance. Since $(\bl, \xi_s)_{\text{opt}} \in S_{ij} $ for some $i,j$ we have $\eta_A((\bl, \xi_s)_{\text{opt}}) = \ba_{ij}$. We then set $\eta_B(\bu) = \bb_{k}$ where $\pa{\bb_k} = \{\ba_{ij} \}$. In a sense $\ba_{ij}$ is the terminal node nearest to $\bu$; having defined $\eta_B$ in such a way forces the parent set of any location to include at least one realization of the process from the same variable. There is no penalty in using $\calD^* = \calD \times \Xi$ as we can write $p(\bw(\bu)\given \pa{\bw(\bu)}) = p(\bw((\bl, \xi_1), \dots, (\bl, \xi_q))\given \pa{\bw(\bu)}) = \prod_{s=1}^q p(w(\bl,\xi_s) \given w(\bl, \xi_1), \dots, w(\bl, \xi_{s-1}), \pa{\bw(\bl)}))$, which also implies that the size of the parent set may depend on the variable index. 
Assumptions of conditional independence across variables can be encoded similarly. Also note that any specific choice of $\eta_B$ induces a partition on $\calU$; let $U_{j} = \{ \bu \in \calU : \eta_B(\bu) = \bb_j \}$, then clearly $\calU = \cup_{j=1}^{m_U} U_j$ with $U_i \cap U_j = \emptyset$ if $i\neq j$. This partition does not necessarily correspond to the partitioning scheme used on $\calS$. $\eta_B$ may by designed to ignore part of the tree and result in $m_U < m_B$. However, we can just drop the unused leaves from $\calG$ and set $\ch{\ba} = \emptyset$ for terminal nodes whose leaf is inactive, resulting in $m_U = m_B$. We will thus henceforth assume that $m_U = m_B$ without loss of generality.

\subsection{\modelnames\ as a standalone spatial process}
We define a valid joint density for any finite set of locations in $\calD^*$ satisfying the Kolmogorov consistency conditions in order to define a valid process. We approach this problem analogously to \cite{nngp} and \cite{meshedgp}. Enumerate each of the $m_S$ reference subsets as $S_i = \{ \bs_{i_1}, \dots, \bs_{i_{n_i}} \}$ where $\{i_1,\ldots,i_{n_i}\}\subset \{1,\ldots,n_{\calS}\}$, and each of the $m_U$ non-reference subsets as $U_i = \{ \bu_{i_1}, \dots, \bu_{i_{n_i}} \}$ where $\{i_1,\ldots,i_{n_i}\}\subset \{1,\ldots,n_{\calU}\}$. Then introduce $\calV = \{ V_1, \dots, V_{m_V} \}$ where $m_V = m_S + m_U$ and $V_i = S_i$ for $i=1, \dots, m_S$, $V_{m_S+i} = U_i$ for $i=1, \dots, m_U$. Then take $\bw_i = (w(\bl_{i_1}), \ldots, w(\bl_{i_{n_i}}))^{\top}$ as the $n_i\times 1$ random vector with elements of $w(\bl)$ for each $\bl\in V_i$. Denote $\bwpa{i} = \bw( \eta^{-1}(\pa{\bv_i}))$. Then 
\begin{equation}\label{eq:pws} 
\begin{aligned}
	\tilde{p}(\bw_{\calS}) = \tilde{p}(\bw_1, \dots, \bw_{m_S})
	= \prod_{r=0}^{M-1} \prod_{i : \{ \bv_i \in \bA_r \} } p(\bw_i \given \bwpa{i}) \qquad & \quad 
	\tilde{p}(\bw_{\calU} \mid \bw_{\calS}) = \prod_{i : \{ \bv_i \in \bB \} } p( \bw_i \given \bwpa{i}) \\
	\tilde{p}(\bw_{\calS}) \tilde{p}(\bw_{\calU} \mid \bw_{\calS}) =  
	\prod_{r=0}^{M-1} \prod_{i : \{ \bv_i \in \bA_r \} } p(\bw_i \given \bwpa{i})
	&\prod_{i : \{ \bv_i \in \bB \} }  p( \bw_i \given \bwpa{i}) 
\end{aligned}
\end{equation}
which is a proper multivariate joint density since $\calG$ is acyclic \citep{lauritzen}.
All locations inside $U_j$ always share the same parent set, but a parent set is not necessarily unique to a single $U_j$. This includes as a special case a scenario in which one can assume 
\begin{align}\label{equation:u_locations_independent}
\tilde{p}(\bw_{\calU} \mid \bw_{\calS}) = \prod_{j=1}^{m_U} \prod_{i=1}^{|U_j|} p( w(\bu_i) \given \bw(\eta^{-1}(\pa{\bb_j})));    
\end{align} 
in this case each location corresponds to a leaf node.
To conclude the construction, for any finite subset of spatial locations $\mathcal{L} \subset \mathcal{D}$ we can let $\calU = \mathcal{L} \setminus \calS$ and obtain 
\[\tilde{p}(\bw_\mathcal{L}) = \int \tilde{p}(\bw_{\calU} \mid \bw_{\calS}) \tilde{p}(\bw_{\calS}) \prod_{ \bs_i \in \calS \setminus \mathcal{L}} d(\bw(\bs_i)),\] 
leading to a well-defined process satisfying the Kolmogorov conditions (see Appendix \ref{appx:kolmogorov}).

\subsubsection{Positioning of spatial locations in conditioning sets} 
In spatial models based on sparse DAGs, larger conditioning sets yield processes that are closer to the base process $p$ in terms of Kullback-Leibler divergence \citep{sudipto_ss20, meshedgp}, denoted as $KL(\cdot \| p)$. The same results cannot be applied directly to \modelnames\ given the treed structure of the DAG. For a given $\calS$, we consider the distinct but related issues of placing individual locations into reference subsets (1) at different levels of the treed hierarchy; (2) within the same level of the hierarchy. 
\begin{proposition}\label{prop:prop1}
Suppose $\calS = \calS_0 \cup \calS_1$ where $S_{0} \cap S_{1} = \emptyset$ and $\calS_1 = S_{11} \cup S_{12}$, $S_{11} \cap S_{12} = \emptyset$. Take $\bs^* \notin \calS$. Consider the graph $\calG = \{ \bV = \{ \bv_0, \bv_1, \bv_2 \}, \bE = \{ \bv_0\to\bv_1, \bv_0\to\bv_2 \} \}$; denote as $p_0$ the density of a \modelname\ using $\eta(\calS_0 \cup \{ \bs^* \}) = \bv_0$, $\eta(S_{11}) = \bv_1$ and $\eta(S_{12}) = \bv_2$, whereas let $p_1$ be the density of a \modelname\ with $\eta(\calS_0) = \bv_0$, $\eta(S_{11} \cup \{ \bs^*\}) = \bv_1$ and $\eta(S_{12}) = \bv_2$. Then $KL(p_1 \| p) - KL(p_0 \| p) > 0.$
\end{proposition}
The proof proceeds by an ``information never hurts'' argument \citep{coverthomas91}.
Denote $\calS^* = \calS \cup \{ \bs^* \}$, $\bw^*= \bw_{\calS^*}$, $w^* = w(\bs^*)$ and $\bw_j^* = (\bw_j^\top, w^*)^\top$. Then
\begin{align*}
    p_0(\bw^*) &= p(\bw_0^*) p(\bw_1 \given \bw_0^*) p(\bw_2 \given \bw_0^*) = p(\bw_0) p(w^* \given \bw_0) p(\bw_1 \given \bw_0, w^*) p(\bw_2 \given \bw_0^*) \\
    p_1(\bw^*) &= p(\bw_0) p(\bw_1^* \given \bw_0) p(\bw_2 \given \bw_0) = p(\bw_0) p(w^* \given \bw_0) p(\bw_1 \given \bw_0, w^*) p(\bw_2 \given \bw_0),
\end{align*}
therefore $p_0(\bw^*) / p_1(\bw^*) = p(\bw_2 \given \bw_0^*) / p(\bw_2 \given \bw_0)$; then by Jensen's inequality
\begin{equation}\label{KL_approx}
\begin{aligned}
KL(p_1 \| p) - KL(p_0 &\| p) = \int \left\{ \log\left(\frac{p(\bw^*)}{p_1(\bw^*)}\right) - \log\left(\frac{p(\bw^*)}{p_0(\bw^*)} \right) \right\} p(\bw^*) d\bw^* \\
&= \int \log\left(\frac{p_0(\bw^*)}{p_1(\bw^*)}\right) p(\bw^*) d\bw^* = \int \log \left(\frac{ p(\bw_2 \given \bw_0^*) }{ p(\bw_2 \given \bw_0) }\right) p(\bw^*) d\bw^*\\
&= \int \log \left(\frac{ p(\bw_2 \given \bw_0^*) }{ p(\bw_2 \given \bw_0) }\right) p(\bw_1, \bw_2, \bw_0^*) d\bw_1 d\bw_2 d\bw_0^* \\
&= \int \left\{ \int \log \left(\frac{ p(\bw_2 \given \bw_0^*) }{ p(\bw_2 \given \bw_0) }\right)  p(\bw_1, \bw_2 \given \bw_0^*)  d\bw_1 d\bw_2 \right\} p(\bw_0^*) d\bw_0^* \geq 0. 
\end{aligned}
\end{equation} 
Intuitively, this shows that there is a penalty associated to positioning reference locations at higher levels of the treed hierarchy. Increasing the size of the reference set at the root augments the conditioning sets at all its children; since this is not true when the increase is at a branch level, the density $p_0$ is closer to $p$ than $p_1$. In other words there is a cost of branching in $\calG$ which must be justified by arguments related to computational efficiency. The above proposition also suggests populating near-root branches with locations of sparsely-observed outcomes. Not doing so in highly imbalanced settings may result in possibly too restrictive spatial conditional independence assumptions. 

\begin{proposition}
Consider the same setup as Proposition \ref{prop:prop1} and let $p_2$ be the density of a \modelname\ such that $\eta(S_{12} \cup \{ \bs^* \}) = \bv_2$. Let $H_p$ be the conditional entropy of base process $p$. Then $H_p(w^* \given \bw_0, \bw_2) < H_p(w^* \given \bw_0, \bw_1)$ implies $KL(p_2 \| p) < KL(p_1 \| p)$.
\end{proposition}
\noindent The density of the new model is \[p_2(\bw^*) = p(\bw_0) p(\bw_1 \given \bw_0) p(\bw_2^* \given \bw_0) = p(\bw_0) p(\bw_1 \given \bw_0) p(\bw_2 \given \bw_0) p(w^* \given \bw_0, \bw_2).\] Then, noting that $p(\bw^*_1 \given \bw_0) = p(\bw_1 \given \bw_0) p(w^* \given \bw_0, \bw_1)$, we get $\frac{p_1(\bw^*)}{p_2(\bw^*)} = \frac{p(w^* \given \bw_0, \bw_1)}{p(w^* \given \bw_0, \bw_2)}$ and
\begin{align*} 
    KL(p_2 \| p) - KL(p_1 &\| p) 
    = \int \log p(w^* \given \bw_0, \bw_1) p(\bw^*) d\bw^* - \int \log p(w^* \given \bw_0, \bw_2) p(\bw^*) d\bw^*\\
    &= H_p(w^* \given \bw_0, \bw_2) - H_p(w^* \given \bw_0, \bw_1).
\end{align*}
This result suggests placing a new reference location $\bs^*$ in the reference subset \textit{least} uncertain about the realization of the process at $\bs^*$. We interpret this as justifying recursive domain partitioning on $\calS$ in spatial contexts in which local spatial clusters of locations are likely less uncertain about process realization in the same spatial region. 
In the remainder of this article, we will consider a given reference set $\calS$ which typically will be based on a subset of observed locations; the combinatorial problem of selecting an optimal $\calS$ (in some sense) is beyond the scope of this article. If $\calS$ is not partitioned, it can be considered as a set of knots or ``sensors'' and one can refer to a large literature on experimental design and optimal sensor placement \citep[see e.g.][and references therein]{krauseetal2008}. It might be possible to extend previous work on adaptive knot placement \citep{pp_adaptive_knots}, but this will come at a steep cost in terms of computational performance.

\section{Bayesian spatial regressions using \modelnames} \label{section:spamtree_gp} 
Suppose we observe an $l$-variate outcome at spatial locations $\bl\in \calD \subset \Re^{d}$ which we wish to model using a spatially-varying regression model:
\begin{equation}\label{eq:linear_svc}
\begin{aligned}
y_j( \bl ) &= \bx_j(\bl)^\top \bbeta_j + \sum_k z_{jk}(\bl) w(\bl, \bxi_k) + \varepsilon_j(\bl), \quad j=1, \dots, l, 
\end{aligned}
\end{equation}
where $y_j(\bl)$ is the $j$-th point-referenced outcome at $\bl$, $\bx_j(\bl)$ is a $p_j \times 1$ vector of spatially referenced predictors linked to constant coefficients $\bbeta_j$, $\varepsilon_j(\bl)\iidsim N(0, \tau^2_j)$ is the measurement error for outcome $j$, and $z_{jk}(\bl)$ is the $k$-th (of $q$) covariates for the $j$-th outcome modeled with spatially-varying coefficient $w(\bl, \bxi_k)$, $\bl \in \calD$, $\bxi_k \in \Xi$.
This coefficient  $w(\bl, \bxi_k)$
corresponds to the $k$-th margin of a $q$-variate Gaussian process $\{\bw(\bl) : \bl \in \calD\}$ denoted as $\bw(\bl) \sim GP(\bzero, \Cov_{\btheta}(\cdot, \cdot))$ with cross-covariance $\Cov_{\btheta}$ indexed by unknown parameters $\btheta$ which we omit in notation for simplicity. A valid cross-covariance function is defined as $\Cov_{\btheta}: \calD\times\calD \rightarrow {\cal M}_{q\times q}$, where ${\cal M}_{q\times q}$ is a subset of the space of all $q\times q$ real matrices $\Re^{q\times q}$. It must satisfy $\Cov(\bl,\bl') = \Cov(\bl',\bl)^{\top}$ for any two locations $\bl, \bl' \in \calD$, and $\sum_{i=1}^n\sum_{j=1}^n \bz_i^{\top}\Cov(\bl_i,\bl_j)\bz_j > 0$ for any integer $n$ and finite collection of points $\{\bl_1,\bl_2,\ldots,\bl_n\}$ and for all $\bz_i \in \Re^{q}\setminus \{\bolds{0}\}$. 

We replace the full GP with a Gaussian \modelname\ for scalable computation considering the $q$-variate multivariate Gaussian process $\bw(\cdot)$ as the base process. Since the $(i,j)$-th entry of $\Cov(\bl,\bl')$ is $\Cov(\bl, \bl')_{i,j}=\mbox{Cov}(w_i(\bl),w_j(\bl'))$, i.e. the covariance between the $i$-th and $j$-th elements of $\bw(\bl)$ at $\bl$ and $\bl'$, we can obtain a covariance function on the augmented domain $\Cov^* : \calD^* \times \calD^* \rightarrow \Re$ as $\Cov^*((\bl, \bxi), (\bl', \bxi')) = \Cov(\bl, \bl')_{i,i'}$ where $\bxi$ and $\bxi'$ are the locations in $\Xi$ of variables $i$ and $j$, respectively. \cite{apanasovich_genton2010} use a similar representation to build valid cross-covariances based on existing univariate covariance functions; their approach amounts to considering $\bxi$ or $\| \bxi - \bxi' \|$ as a parameter to be estimated. Our approach can be based on any valid cross-covariance as we may just set $\Xi = \{1, \dots, q\}$. Refer to e.g. \cite{genton_ccov} for an extensive review of cross-covariance functions for multivariate processes. Moving forward, we will not distinguish between $\Cov^*$ and $\Cov$. The linear multivariate spatially-varying regression model (\ref{eq:linear_svc}) allows the $l$ outcomes to be observed at different locations; we later consider the case $l=q$ and $\bZ(\bl) = I_q$ resulting in a multivariate space-varying intercept model. 

\subsection{Gaussian \modelnames}
Enumerate the set of nodes as $\bV = \{ \bv_1, \dots, \bv_{m_V}\}$, $m_V = m_S + m_U$ and denote $\bw_i = w(\eta^{-1}(\bv_i))$, $\Cov_{ij}$ as the $n_i \times n_j$ covariance matrix between $\bw_i$ and $\bw_j$, $\Cov_{i,[i]}$ the $n_i\times J_i$ covariance matrix between $\bw_i$ and $\bwpa{i}$, $\Cov_{i}$ the $n_i \times n_i $ covariance matrix between $\bw_{i}$ and itself, and $\Cov_{[i]}$ the $J_i \times J_i $ covariance matrix between $\bwpa{i}$ and itself. 
A base Gaussian process induces $\tp(\bw_{\calS}) = \prod_{j : \{\bv_j \in \bA \}} N( \bw_j \mid \bH_{j} \bwpa{j}, \bR_{j})$, where 
\begin{equation}\label{equation:h_and_r} 
	\bH_{j} = \Cov_{j, [j]} \Cov^{-1}_{[j]} \quad\text{and}\quad \bR_{j} = \Cov_{j} - \Cov_{j, [j]}\Cov^{-1}_{[j]} \Cov_{[j],j},
\end{equation}
implying that the joint density $\tilde{p}(\bw_{\calS})$ is multivariate normal with covariance $\tilde{\Cov}_{\calS}$ and precision matrix $\tilde{\Cov}^{-1}_{\calS}$. At $\calU$ we have $\tp( \bw_{\calU} \mid \bw_{\calS}) = \prod_{j : \{ \bv_j \in \bB \}} N( \bw_{j} \mid \bH_{j}\bwpa{j}, \bR_{j} )$, where $\bH_{j}$ and $\bR_{j}$ are as in (\ref{equation:h_and_r}). All quantities can be computed using the base cross-covariance function. Given that the $\tp$ densities are Gaussian, so will be the finite dimensional distributions.

The treed graph $\calG$ leads to properties which we analyze in more detail in Appendix \ref{appendix:gp_properties} and summarize here. 
For two nodes $\bv_i, \bv_j \in \bV$ denote the \textit{common descendants} as $\text{cd}(\bv_i, \bv_j) = (\{ \bv_i \} \cup \ch{\bv_i}) \cap (\{ \bv_j \} \cup \ch{\bv_j})$. If $\bv_i \in \pa{\bv_j}$ denote $\bH_{i\to j}$ and $\bH_{\setminus i \to j}$ as the matrix obtained by subsetting $\bH_j$ to columns corresponding to $\bv_i$, or to $\pa{\bv_j}\setminus\{\bv_i \}$, respectively. Similarly define $\bw_{[i\to j]} = \bw_i$ and $\bw_{[\setminus i \to j]}$. As a special case, if the tree depth is $\delta=1$ and $\{ \bv_j\} = \pa{\bv_i}$ then $\text{cd}(\bv_i, \bv_j) = \{ \bv_i \}$, $\bH_{i\to j} = \bH_j$, and $\bw_{[i\to j]} = \bwpa{j}$.
Define $\bbH$ as the matrix whose $(i,j)$ block is $\bbH_{ij} = \bO_{n_i \times n_j}$ if $\bv_j \notin \pa{\bv_i}$, and otherwise $\bbH_{ij} = \bH_{j\to i}$. 

\subsubsection{Precision matrix}
The $(i,j)$ block of the precision matrix at both reference and non-reference locations $\tilde{\Cov}^{-1}$ is denoted by $\tilde{\Cov}^{-1}(i, j)$, with $i,j=1, \dots, m_V$ corresponding to nodes $\bv_i, \bv_j \in \bV$ for some $i,j$; it is nonzero if $\text{cd}(\bv_i, \bv_j) = \emptyset$, otherwise:
\begin{equation}\label{equation:precision_blocks_main}
\begin{aligned}
\tilde{\Cov}^{-1}(i, j) &= \sum\limits_{\bv_k \in \text{cd}(\bv_i, \bv_j)}
    (\bI_{ki} - \bH_{i\to k})^\top \bR_k^{-1} (\bI_{kj} - \bH_{j\to k})\\
    &= \sum\limits_{\bv_k \in \text{cd}(\bv_i, \bv_j)}
    (\bI_{ki} - \bbH_{ki})^\top \bR_k^{-1} (\bI_{kj} - \bbH_{kj}),
\end{aligned}
\end{equation}
where $\bI_{ij}$ is the $(i,j)$ block of an identity matrix with $n_{\calS}+n_{\calU}$ rows and is nonzero if and only if $i=j$. We thus obtain that the number of nonzero elements of $\tilde{\Cov}^{-1}$ is 
\begin{align} \label{equation:sparsity}
\text{nnz}(\tilde{\Cov}^{-1}) &= \sum_{i=1}^{m_V} \left( 2n_{i} J_{i} + n_{i}^2\bolds{1}\{\bv_i \in \bV\}\right),
\end{align}
where $n_{i} = |\eta^{-1}(\bv_{i})|$, $J_{i} = | \eta^{-1}(\pa{\bv_{i}})|$, and by symmetry $(\tilde{\Cov}^{-1}(i,j))^{\top} = \tilde{\Cov}^{-1}(j,i)$. 

If $\delta > 1$, the size of $\Cov_{[i]}$ is larger for nodes $\bv_i$ at levels of the treed hierarchy farther from $\bA_{M_{\delta}}$. However suppose $\bv_i, \bv_j$ are such that $\pa{\bv_j} = \{ \bv_i \} \cup \pa{\bv_i}$. Then computing $\Cov_{[j]}^{-1}$ proceeds more cheaply by recursively applying the following:
\begin{align} \label{equation:nested_inverse_main}
    \Cov_{[j]}^{-1} = \begin{bmatrix} \Cov_{[i]}^{-1} + \bH_i^\top \bR_i^{-1}\bH_i & - \bH_i^\top \bR_i^{-1} \\
    -\bR_i^{-1}\bH_i  & \bR_i^{-1}\end{bmatrix}.
\end{align}

\subsubsection{Induced covariance}
Define a path from $\bv_k$ to $\bv_j$ as $\calP_{k \to j} = \{\bv_{i_{1}}, \dots, \bv_{i_{r}} \}$ where $\bv_{i_{1}} = \bv_k$, $\bv_{i_{r}} = \bv_j$, and  $\bv_{i_h} \in \pa{\bv_{i_{h+1}}}$. The longest path $\tilde{\calP}_{k \to j}$ is such that if $\bv_k \in \bA_{r_k}$ and $\bv_j \in \bA_{r_j}$ then $|\tilde{\calP}_{k \to j}| = r_j - r_k + 1$. The shortest path $\bar{\calP}_{k \to j}$ is the path from $\bv_k$ to $\bv_j$ with minimum number of steps. We denote the longest path from the root to $\bv_j$ as $\tilde{\calP}_{0\to j}$; this corresponds to the full set of ancestors of $\bv_j$, and $\pa{\bv_j} \subset \tilde{\calP}_{0\to j}$. For two nodes $\bv_i$ and $\bv_j$ we have $(\pa{\bv_i} \cap \pa{\bv_j}) \subset (\tilde{\calP}_{0\to i} \cap \tilde{\calP}_{0\to j})$. We define the \textit{concestor} between $\bv_i$ and $\bv_j$ as $\text{con}(\bv_i, \bv_j) = \arg \max_{\bv_k \in \bV} \{ k : \calP_{k\to i} \cap \calP_{k\to j} \neq \emptyset \} $ i.e. the last common ancestor of the two nodes. 

Take the path $\tilde{\calP}_{M_{\delta} \to j}$ in $\calG$ from a node at $\bA_{M_{\delta}}$ leading to $\bv_j$. After defining the cross-covariance function $\bK_i(\bl, \bl') = \Cov_{\bl, \bl'} - \Cov_{\bl, [i]} \Cov^{-1}_{[i]} \Cov_{[i], \bl'}$ and denoting $\bK_i(\bl, s) = \bK_i(\bl, \eta^{-1}(\bv_s))$ we can write 
\begin{equation}\label{equation:additive_repres_main}
    \begin{aligned}
        \bw_j = \sum\limits_{s=i_{M_{\delta}}}^{i_{r-1}} \bK_{s}(j,s) \bK_{s}^{-1}(s,s) \be_s + \be_j,
    \end{aligned}
\end{equation}
where for $s>i_{M_{\delta}}$ the $\be_s$ are independent zero-mean GPs with covariance $K_s(\bl, \bl')$ and we set $K_{i_{M_{\delta}}}(\bl, \bl') = \Cov(\bl, \bl')$ and  $\be_{i_{M_{\delta}}} = \bw_{i_{M_{\delta}}} \sim N(0, \Cov_{i_{M_{\delta}}})$. 
Take two locations $\bl, \bl'$ such that $\bv_i = \eta(\bl), \bv_j = \eta(\bl')$ and let $\bv_z = \text{con}(\bv_i, \bv_j)$; if $\pa{\bv_i} \cap \pa{\bv_j} \neq \emptyset$ then the above leads to
\begin{align}\label{equation:cov_recursive}
    Cov_{\tp}(\bw(\bl), \bw(\bl')) &= \sum\limits_{s \in \pa{\bv_i} \cap \pa{\bv_j}} \bK_{s}(\bl,s) \bK_{s}^{-1}(s,s)\bK_{s}(s,\bl') + \bolds{1}\{\bl = \bl'\}\bK_j(\bl, \bl'),
\end{align}
where $\bK_z(\bl, \bl') = \Cov(\bl, \bl')$. If $\pa{\bv_i} \cap \pa{\bv_j} = \emptyset$ take the shortest paths $\bar{\calP}_{z \to i} = \{ i_1, \dots, i_{r_i}\}$ and $\bar{\calP}_{z\to j} = \{ j_1, \dots, j_{r_j} \}$; setting $\bF_{i_h} = \Cov_{i_h, i_{h-1}} \Cov_{i_{h-1}}^{-1}$ we get
\begin{align}\label{equation:cov_limited}
    Cov_{\tp}(\bw(\bl), \bw(\bl')) &= \bF_{i_{r_i}} \cdots \bF_{i_{1}} \Cov_{z} \bF_{j_{1}}^{\top} \cdots \bF_{j_{r_j}}^{\top}.
\end{align}
In particular if $\delta=M$ then $\pa{\bv_i} \cap \pa{\bv_j} \neq \emptyset$ for all $i,j$ and only (\ref{equation:cov_recursive}) is used, whereas if $\delta=1$ then the only scenario in which (\ref{equation:cov_recursive}) holds is $\{ \bv_z \} = \pa{\bv_i} \cap \pa{\bv_j}$ in which case  the two are equivalent. 
In univariate settings, the special case in which $\delta = M$, and hence $M_{\delta} = 0$, leads to an interpretation of (\ref{equation:additive_repres_main}) as a basis function decomposition; considering  all leaf paths $\calP_j$ for $\bv_j \in \bB$, this leads to an MRA \citep{katzfuss_jasa17, katzfussgong2019}. On the other hand, keeping other parameters constant, $\delta<M$ and in particular $\delta=1$ may be associated to savings in computing cost, leading to a trade-off between graph complexity and size of reference subsets; see Appendix \ref{appendix:computing_cost}.

\subsubsection{Block-sparse Cholesky decompositions}
In recent work \cite{jurekkatzfuss2020} consider sparse Cholesky decompositions of covariance and precision matrices for treed graphs corresponding to the case $\delta=M$ above in the context of space-time filtering; their methods involve sparse Cholesky routines on reverse orderings of $\tilde{\Cov}^{-1}$ at the level of individual locations. In doing so, the relationship between Cholesky decompositions and $\calG$, $\tilde{\Cov}^{-1}$ and the block structure in $\calS$ remains somewhat hidden, and sparse Cholesky libraries are typically associated to bottlenecks in MCMC algorithms. However we note that a consequence of (\ref{equation:precision_blocks_main}) is that it leads to a direct algorithm, for any $\delta$, for the block-decomposition of any symmetric positive-definite matrix $\bLambda$ conforming to $\calG$, i.e. with the same block-sparse structure as $\tilde{\Cov}^{-1}$. This allows us to write $\bLambda = (\bI - \bL)^\top \bD (\bI - \bL)$ where $\bI$ is the identity matrix, $\bL$ is block lower triangular with the same block-sparsity pattern as $\bbH$ above, and $\bD$ is block diagonal symmetric positive-definite. In Appendix \ref{appendix:precision_decomposition} we outline Algorithm \ref{algorithm:precision_decomposition} which \textit{(i)} makes direct use of the structure of $\calG$,  \textit{(ii)} computes the decomposition at blocks of reference and non-reference locations, and \textit{(iii)} requires no external sparse matrix library, in particular no sparse Cholesky solvers. Along with Algorithm \ref{algorithm:precision_decomp_invtri} for the block-computation of $(\bI-\bL)^{-1}$, it can be used to compute $\bLambda^{-1}= (\tilde{\Cov}^{-1} + \bSigma)^{-1}$ where $\bSigma$ is a block-diagonal matrix; it is thus useful in computing the Gaussian integrated likelihood.

\subsection{Estimation and prediction} \label{section:general:estimation}
We introduce notation to aid in obtaining the full conditional distributions. Write (\ref{eq:linear_svc}) as
\begin{equation}\label{eq:linear_svc_exp}
\begin{aligned}
\by(\bl) &= \bX(\bl) \bbeta + \bZ(\bl) \bw(\bl) + \beps(\bl),
\end{aligned}
\end{equation}
where $\by(\bl) = ( \{y_j(\bl)\}_{j=1}^l )^{\top}$, $\beps(\bl) = ( \{\varepsilon_j(\bl)\}_{j=1}^l )^{\top} \sim N(\bzero, \bD_{\tau})$, $\bD_{\tau} = \text{diag}(\tau^2_1, \dots, \tau^2_l)$, $\bX(\bl) = \bdiag\{ \bx_j(\bl)^{\top}, j=1,\dots,l \}$, $\bbeta = (\bbeta_{p_1}^\top, \dots, \bbeta_{p_j}^{\top})^{\top}$. The $l \times q$ matrix $\bZ(\bl) = (\bz_j(\bl)^{\top}, j=1,\dots,l)$ with $\bz_j(\bl)^{\top} = (z_{jk}(\bl), k=1, \dots,q)$ acts a design matrix for spatial location $\bl$. Collecting all locations along the $j$-th margin, we build $\calT_j = \{ \bl_1^{(j)}, \dots, \bl_{N_j}^{(j)}\}$ and $\calT = \cup_j \calT_j$. We then call $\by^{(j)} = (y_j(\bl_1^{(j)}), \dots, y_j(\bl_{N_j}^{(j)}))^\top$ and $\beps^{(j)}$ similarly, $\bX^{(j)}= (\bx_j(\bl_1^{(j)}), \dots, \bx_j(\bl_{N_j}^{(j)}))^\top$, $\bw^{(j)}= (\bw(\bl_1^{(j)}, \bxi)^{\top}, \dots, \bw(\bl_{N_j}^{(j)}, \bxi)^{\top})^\top$ and $\bZ^{(j)} = \bdiag\{ \bz_{j}(\bl_s^{(j)})^\top \}_{s=1}^{N_j}$. The full observed data are $\by, \bX, \bZ$. Denoting the number of observations as $n = \sum_{j=1}^l N_j$, $\bZ$ is thus a $n \times q n$ block-diagonal matrix, and similarly $\bw$ is a $qn \times 1$ vector. We introduce the diagonal matrix $\bD_n$ such that $\text{diag}(\bD_n) = (\tau^2_1\bolds{1}_{N_1}^{\top}, \dots, \tau^2_l \bolds{1}_{N_l}^{\top})^{\top}$.

By construction we may have $\eta(S_i) = \bv_i$ and $\eta(S_j) = \bv_j$ such that $(\bl, \bxi) \in S_i$ and $(\bl', \bxi') \in S_j$ where $\bl'=\bl$, $\bxi\neq \bxi'$ and similarly for non-reference subsets. Suppose $\calA \subset \calD \times \Xi$ is a generic reference or non-reference subset. We denote $\bar{\calA} \subset \calD \times \Xi$ as the set of all combinations of spatial locations of $\calA$ and variables i.e. $\bar{\calA} = \calA\big|_{\calD} \times \calA\big|_{\Xi}$ where $\calA\big|_{\calD} \subset \calD$ is the set of unique spatial locations in $\calA$ and $\calA\big|_{\Xi}$ are the unique latent variable coordinates. 
By subtraction we find $\calA_{-} = \bar{\calA} \setminus \calA$ as the set of locations whose spatial location is in $\calA$ but whose variable is not. Let $\by(\bar{\calA}) = \by(\calA) = (\{ \by(\bl), \bl \in \calA\big|_{\calD} \} )^\top$,  $\bX(\bar{\calA}) = \bX(\calA) = \bdiag\{ \bX(\bl)^{\top}, \bl \in \calA\big|_{\calD}\}$; values corresponding to unobserved locations will be dealt with by defining $\tilde{\bD}_n(\calA)$ as the diagonal matrix obtained from $\bD_n$ by replacing unobserved outcomes with zeros. 
Denote $\bZ(\bar{\calA}) = \bdiag\{ \bZ(\bl), \bl \in \calA\big|_{\calD} \}$ and  $\bw(\bar{\calA})$ similarly. If $\calA$ includes $L$ unique spatial locations then $\by(\bar{\calA})$ is a $L\ l\times 1$ vector and $\bX(\calA)$ is a $L\ l \times pl$ matrix. In particular, $\bZ(\bar{\calA})$ is a $L\ l \times Lql$ matrix; the subset of its columns with locations in $\calA$ is denoted as $\bZ(\calA)$ whereas at other locations we get $\bZ(\calA_{-})$. We can then separate the contribution of $\bw(\calA)$ to $\by({\calA})$ from the contribution of $\bw(\calA_{-})$ by writing $\by({\calA}) = \bX({\calA}) \bbeta + \bZ(\calA_{-}) \bw(\calA_{-}) + \bZ(\calA) \bw(\calA) + \beps({\calA})$, using which we let $\tilde{\by}({\calA}) = \by({\calA}) - \bX({\calA}) \bbeta - \bZ(\calA_{-}) \bw(\calA_{-})$.

With customary prior distributions $\bbeta \sim N(\bzero, \bV_{\beta})$ and $\tau^2_j \sim Inv.Gamma(a_{\tau}, b_{\tau})$ along with a Gaussian \modelname\ prior on $\bw$, we obtain the posterior distribution as
\begin{align} \label{equation:posterior_distribution}
    p(\bw, \bbeta, \{\tau_j^2\}_{j=1}^l, \btheta \given \by) &\propto p(\by \given \bw, \bbeta, \{\tau_j^2\}_{j=1}^l) p(\bw \given \btheta) p(\btheta) p(\bbeta) \prod_{j=1}^l p(\tau^2_j).  
\end{align}
We compute the full conditional distributions of unknowns in the model, save for $\btheta$; iterating sampling from each of these distributions corresponds to a Gibbs sampler which ultimately leads to samples from the posterior distribution above.

\subsubsection{Full conditional distributions}
 The full conditional distribution for $\bbeta$ is Gaussian with covariance $\bSigma^*_{\bbeta} = (\bV_{\bbeta}^{-1} + \bX^{\top} \bD_n^{-1} \bX)^{-1}$ and mean $\mu^*_{\bbeta} = \bSigma_{\bbeta} \bX^{\top} \bD_n^{-1} (\by - \bZ \bw)$. For $j=1, \dots, l$, $p(\tau^2_j \given \bbeta, \by, \bw) = Inv.Gamma(a_{\tau, j}^*, b^*_{\tau, j})$ where $a_{\tau, j}^* = a_{\tau} + N_j/2$ and $b^*_{\tau, j} = b_{\tau} + \frac{1}{2}\bE^{(j)\top}\bE^{(j)}$ with $\bE^{(j)} = \by^{(j)} - \bX^{(j)}\bbeta_j - \bZ^{(j)}\bw^{(j)}$.

Take a node $\bv_i \in \bV$. If $\bv_i \in \bA$ then $\eta^{-1}(\bv_i) = S_i$ and for $\bv_j \in \ch{\bv_i}$ denote $\tilde{\bw}_j = \bw_j - \bH_{\setminus i \to j} \bw_{[\setminus i \to j]}$. The full conditional distribution of $\bw_i$ is $N(\bmu_{i}, \bSigma_{i})$, where
\begin{equation}\label{equation:reference_full_conditional}
\begin{aligned}
\bSigma_{i}^{-1} = \bZ(S_i)^{\top}\bD_n(S_i)^{-1} & \bZ(S_i) + \bR_i^{-1} + \bF^{(c)}_i \\
\bSigma_{i}^{-1}\bmu_{i} = \bZ(S_i)^{\top} \bD_n(S_i)^{-1} & \tilde{\by}(S_i) + \bR_i^{-1}\bH_i\bw_{[i]} + \bm^{(c)}_i\\
\bF^{(c)}_i = \sum_{j: \{\bv_j \in \ch{\bv_i}\}} \bH_{i\to j}^{\top} \bR_j^{-1} \bH_{i\to j} \qquad & \qquad \bm^{(c)}_i = \sum_{j: \{\bv_j \in \ch{\bv_i}\}} \bH_{i\to j}^{\top} \bR_j^{-1} \tilde{\bw}_j
\end{aligned}
\end{equation}
If $\bv_i \in \bB$ instead $\bSigma_{i} = (\bZ(U_i)^{\top}\bD_n(U_i)^{-1} \bZ(U_i) + \bR_i)^{-1}$ and $\bmu_{i} = \bSigma_i (\bZ(U_i)^{\top} \bD_n(U_i)^{-1}\tilde{\by}(U_i) + \bR_i^{-1}\bH_i\bw_{[i]})$. Sampling of $\bw$ at nodes at the same level $r$ proceeds in parallel given the assumed conditional independence structure in $\calG$. It is thus essential to minimize the computational burden at levels with a small number of nodes to avoid bottlenecks. In particular computing $\bF^{(c)}_i$ and $\bm^{(c)}_i$ can become expensive at the root when the number of children is very large. In Algorithm \ref{algorithm:gibbs} we show that one can efficiently sample at a near-root node $\bv_i$ by updating $\bF^{(c)}_i$ and $\bm^{(c)}_i$ via message-passing from the children of $\bv_i$. 

\begin{figure}[tp]
\vspace*{-\baselineskip}
\begin{minipage}{\columnwidth}
\begin{algorithm}[H]
\small
\SetKwInOut{Input}{Input}
\textbf{Initialize:} \( \text{\Large$\ell $} = 0 \)\;
 \For{$r \in \{0, \dots, M\}$}{
    \For(\tcp*[f]{[parallel for]}){$j : \{ \bv_{j} \in \bV_r\}$}{
        Compute $\bR_j^{-1} = (\Cov_j - \Cov_{j,[j]} \Cov_{[j]}^{-1} \Cov_{[j],j})^{-1}$ and $|\bR_j^{-1}|$\; 
        $\text{\Large$\ell $} = \text{\Large$\ell $} + \frac{1}{2}\log|\bR_j^{-1}| - \frac{1}{2} (\bw_j - \bH_j \bw_{[j]})^{\top} \bR_j^{-1} (\bw_j - \bH_j \bw_{[j]})$\;
        \If{$\ch{\bv_j}\neq \emptyset $}{
            Identify $\bv_i \in \ch{\bv_j}$ such that $\bv_i \in \bV_{r+1}$\;
            Compute and store $\Cov_{[i]}^{-1}$ (possibly via (\ref{equation:nested_inverse_main}))\;
        }
    }
}
\textbf{Result:} \( \exp(\text{\Large$\ell $}) \propto p(\bw \given \btheta) = \prod_i N(\bw_i \given \bH_i \bw_{[i]}, \bR_i) \).
\caption{Computing $p(\bw \given \btheta)$.}\label{algorithm:wpriorcompute}
\normalsize
\end{algorithm}
\end{minipage}
\begin{minipage}{\columnwidth}
\begin{algorithm}[H]
\small
\SetKwInOut{Input}{Input}
\textbf{Input:}$\Cov_{[j]}$ for all $j$ from Algorithm \ref{algorithm:wpriorcompute};\\
$\bW_e = \bigcup\limits_{r \text{ is even}} \bV_r$; $\bW_o = \bigcup\limits_{r \text{ is odd}} \bV_r$;\\
 \For{$i \in \{e, o\}$}{
    \For(\tcp*[f]{[parallel for]}){$j : \{ \bv_{j} \in \bW_i\}$}{
        Sample $\bw_j \sim N(\bmu_j, \bSigma_j)$ using (\ref{equation:reference_full_conditional})\;
        Let $\pa{\bv_{j}}=\{ \bv_{p} \}$, then $\bm^{(c)}_p = \bH^{\top}_{j} \bR^{-1}_j \bw_j$ and 
        $\bF^{(c)}_p = \bH^{\top}_{j} \bR^{-1}_j \bH_{j}$\;
    }
  }
 \textbf{Result:} sample from $p(\bw_j \given \bw_{-j},  \by, \bbeta, \btheta, \bolds{\tau})$ for all $\bv_j \in \bV$.
 \caption{Sampling from the full conditional distribution of $\bw_i$ when $\delta = 1$.}\label{algorithm:gibbs_limited}
 \normalsize
\end{algorithm}
\end{minipage}
\begin{minipage}{\columnwidth}
\begin{algorithm}[H]
\small
\SetKwInOut{Input}{Input}
\Input{$\Cov_{[j]}$ for all $j$ from Algorithm \ref{algorithm:wpriorcompute}}
\textbf{Initialize:} for all $i$, $\bolds{m}^{(c)}_i = \bzero_{n_i \times 1}$ and $\bF^{(c)}_i = \bO_{n_i \times n_i}$\;
 \For{$r \in \{M, \dots, 0\}$}{
    \For(\tcp*[f]{[parallel for]}){$j : \{ \bv_{j} \in \bV_r\}$}{
        Sample $\bw_j \sim N(\bmu_j, \bSigma_j)$ using (\ref{equation:reference_full_conditional})\;
        \For{$p : \{ \bv_{p} \in \pa{\bv_{j}} \}  $}{
            $\bm^{(c)}_p = \bm^{(c)}_p + \bH^{\top}_{p \to j} \bR^{-1}_j \bw_j$\;
            $\bF^{(c)}_p = \bF^{(c)}_p + \bH^{\top}_{p \to j} \bR^{-1}_j \bH_{p \to j}$\;
        }
    }
  }
 \textbf{Result:} sample from $p(\bw_j \given \bw_{-j},  \by, \bbeta, \btheta, \bolds{\tau})$ for all $\bv_j \in \bV$.
 \caption{Sampling from the full conditional distribution of $\bw_j$ when $\delta = M$.}\label{algorithm:gibbs}
 \normalsize
\end{algorithm}

\end{minipage}

\end{figure}

\subsubsection{Update of $\btheta$}
The full conditional distribution of $\btheta$---which may include $\bxi_j$ for $j=1,\dots, q$ or equivalently $\delta_{ij} = \| \bxi_i - \bxi_j\|$ if the chosen cross-covariance function is defined on a latent domain of variables---is not available in closed form and sampling a posteriori can proceed via Metropolis-Hastings steps which involve accept/reject steps with acceptance probability $\alpha = \min \{1, \frac{p(\bw \given \btheta') p(\btheta') q(\btheta \given \btheta')}{p(\bw \given \btheta) p(\btheta) q(\btheta' \given \btheta)}\} $. In our implementation, we adaptively tune the standard deviation of the proposal distribution via the robust adaptive Metropolis algorithm \citep[RAM;][]{vihola2012}. In these settings, unlike similar models based on DAG representations such as NNGPs and MGPs, direct computation via $p(\bw \given \btheta) = \prod_i N(\bw_i \given \bH_i \bw_{[i]}, \bR_i)$ is inefficient as it requires computing $\Cov^{-1}_{[i]}$ whose size grows along the hierarchy in $\calG$. We thus outline Algorithm \ref{algorithm:wpriorcompute} for computing $p(\bw \given \btheta)$ via (\ref{equation:nested_inverse_main}). As an alternative we can perform the update using ratios of $p(\by \given \bbeta, \btheta, \bolds{\tau}) = \int p(\by \given \bw, \bbeta, \bolds{\tau}) p(\bw \given \btheta) d\bw = N(\by \given \bX \bbeta, \bZ \tilde{\Cov} \bZ^{\top} + \bD_n)$ using Algorithms \ref{algorithm:precision_decomposition} and \ref{algorithm:precision_decomp_invtri} outlined in Appendix \ref{appendix:precision_decomposition} which require no sparse matrix library. 

\subsubsection{Graph coloring for parallel sampling}
An advantage of the treed structure of $\calG$ is that it leads to fixed graph coloring associated to parallel Gibbs sampling; no graph coloring algorithms are necessary \citep[see e.g.][]{molloyreed2002, lewis2016}. Specifically, if $\delta=M$ (full depth) then there is a one to one correspondence between the $M+1$ levels of $\calG$ and graph colors, as evidenced by the parallel blocks in Algorithms \ref{algorithm:wpriorcompute} and \ref{algorithm:gibbs}. In the case $\delta=1$, $\calG$ is associated to only two colors alternating the odd levels with the even ones. This is possible because the Markov blanket of each node at level $r$, with $r$ even, only includes nodes at odd levels, and vice-versa.

\subsubsection{Prediction of the outcome at new locations}
The Gibbs sampling algorithm will iterate across the above steps and, upon convergence, will produce samples from $p(\bbeta, \{ \tau^2_j \}_{j=1}^q, \bw \mid \by)$. We obtain posterior predictive inference at arbitrary $\bl \in \calD$ by evaluating $p(\by(\bl)\given \by)$. If $\bl \in \calS \cup \calU$, then we draw one sample of $\by(\bl) \sim N(\bX(\bl)^\top \bbeta + \bZ(\bl)^\top \bw(\bl), \bD_n(\bl))$ for each draw of the parameters from $p(\bbeta, \{ \tau^2_j \}_{j=1}^l, \bw \mid \by)$. Otherwise, considering that $\eta(\bl) = \bv_j \in \bB$ for some $j$, with parent nodes $\pa{\bv_j}$, we sample $\bw(\bl)$ from the full conditional $N(\bmu_{\bl}^*, \bSigma_{\bl}^*)$, where $\bSigma_{\bl}^* = (\bZ(\bl)\bD_n(\bl)^{-1} \bZ(\bl)^{\top} + \bR_{\bl}^{-1})^{-1}$ and $\bmu_{\bl}^* = \bSigma_{\bl}^*(\bZ(\bl) \bD^{-1} (\by(\bl) - \bX(\bl)^\top \bbeta ) + \bR_{\bl}^{-1} \bH_{\bl} \bwpa{j})$, then draw $\by(\bl) \sim N(\bX(\bl)^\top \bbeta + \bZ(\bl)^\top \bw(\bl), \bD_n)$.

\subsubsection{Computing and storage cost}\label{section:compute_cost}
The update of $\tau_j^2$ and $\bbeta$ can be performed at a minimal cost as typically $p = \sum_{j=1}^l p_j$ is small; almost all the computation budget must be dedicated to computing $p(\bw \given \btheta)$ and sampling $p(\bw \given \by, \bbeta, \bolds{\tau}^2)$. Assume that reference locations are all observed $\calS \subset \calT$ and that all reference subsets have the same size i.e. $|S_i| = N_s$ for all $i$. We show in Appendix \ref{appendix:computing_cost} that the cost of computing \modelnames\ is $O(n N_s^2)$. As a result, \modelnames\ compare favorably to other models specifically in not scaling with the cube of the number of samples. $\delta$ does not impact the computational order, however, compared to $\delta=M$, choosing $\delta=1$ lowers the cost by a factor of $M$ or more. For a fixed reference set partition and corresponding nodes, choosing larger $\delta$ will result in stronger dependence between leaf nodes and nodes closer to the root, and this typically corresponds to leaf nodes being assigned conditioning sets that span larger distances in space. The computational speedup corresponding to choosing $\delta=1$ can effectively be traded for a coarser partitioning of $\calS$, resulting in large conditioning sets that are more local to the leaves.

\section{Applications} \label{section:applications}
We consider Gaussian \modelnames\ for the multivariate regression model (\ref{eq:linear_svc_exp}). Consider the spatial locations $\bl, \bl' \in \calD$ and the locations of variables $i$ and $j$ in the latent domain of variables $\bxi_i, \bxi_j \in \Xi$, then denote $\bh = \| \bl - \bl' \|$, $\Delta = \delta_{ij} = \| \bxi_i - \bxi_j \|$, and \begin{align*}
	C(\bh, \Delta) = \frac{\exp\left\{ - \phi \| \bh \|/\exp\left\{\frac{1}{2} \beta \log(1+\alpha \Delta )\right\} \right\} }{ \exp\left\{ \beta \log(1+\alpha \Delta ) \right\}}.
\end{align*}
For $j=1, \dots, q$ we also introduce $C_j(\bh) = \exp\left\{ - \phi_j \| \bh \|\right\}$. A non-separable cross-covariance function for a multivariate process can be defined as
\begin{equation}\label{eq:apanasovich_genton_covariance2}
    \begin{aligned}
    \text{Cov}(w(\bl, \bxi_i), w(\bl', \bxi_j)) = \Cov_{ij}(\bh) &= \begin{cases}
    \sigmasq_{i1} C(\bh, \delta_{ij}) + \sigmasq_{i2} C_i(\bh) & \text{if } i=j \\
    \sigma_{i1} \sigma_{j1} C(\bh, \delta_{ij}) & \text{if } i\neq j,
    \end{cases}
    \end{aligned}
\end{equation}
which is derived from eq. (7) of \cite{apanasovich_genton2010}; locations of variables in the latent domain are unknown, therefore $\btheta = \{\sigma_{i1}, \sigma_{i2}, \phi_i\}_{i=1,\dots,q} \cup \{ \delta_{ij} \}_{i=1,\dots, q}^{j<i} \cup \{ \alpha, \beta, \phi\}$ for a total of $3q + q(q-1)/2 + 3$ unknown parameters.

\subsection{Synthetic data} \label{section:simulations}
In this section we focus on bivariate outcomes ($q=2$) using the cross-covariance (\ref{eq:apanasovich_genton_covariance2}). For simplicity we set $\delta_{21} = 1, \alpha=1, \beta=1$. For each combination of $\sigma_{ij} \in \{1, 2\}$ for $i,j \in \{1,2\}$, $\phi_1 = \phi_2 \in \{0.1, 1, 10\}$, and $\phi \in \{ 0.1, 1, 10 \}$ we generate 25 data sets. Considering model (\ref{eq:linear_svc_exp}), we set $\bbeta = \bzero$, $\bZ = I_q$ and take the sampling locations as a regular grid of size $70\times70$ for a total of 4,900 spatial locations.
\begin{figure}
	\centering 
		\includegraphics[width=.95\textwidth]{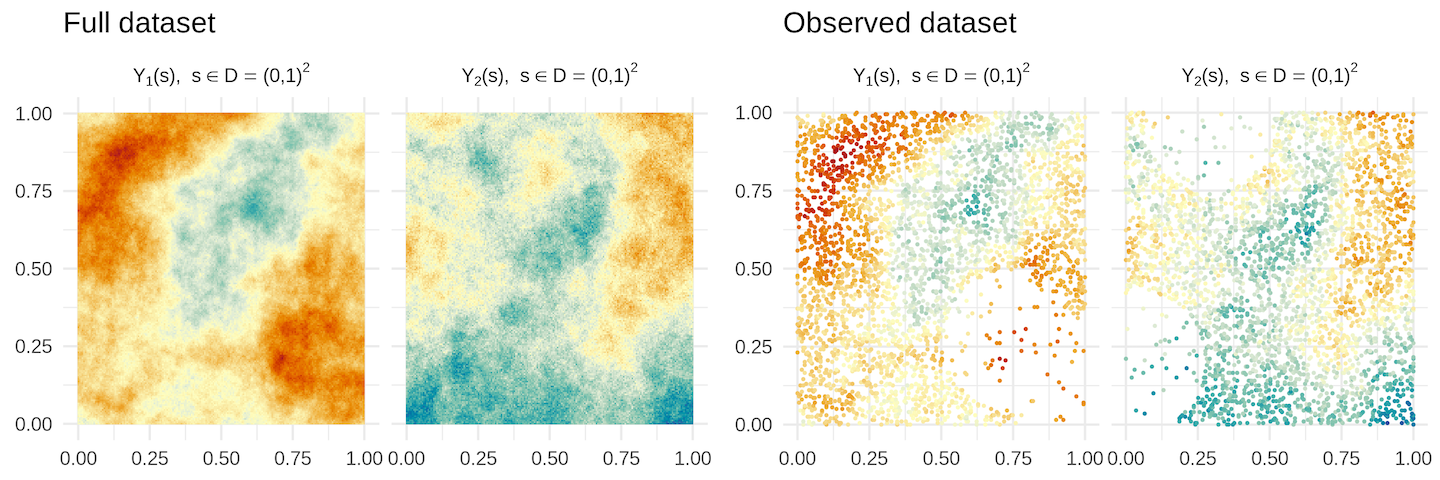}
		\caption{Left half: \textit{Full data set} -- a bivariate outcome is generated on 4,900 spatial locations. Right half: \textit{Observed data set} -- the training sample is built via subsampling each outcome at a smaller set of locations.} 
	\label{figure:synthetic_example} 
\end{figure}
We simulate the spatial effects by sampling the full GP; the nuggets for the two outcomes are set to $\tau^2_1 = 0.01$ and $\tau^2_{2}=0.1$. As a result, we obtain 900 data sets, each simulating a realization of a bivariate outcome. We mimick real-world data measured irregularly in space by replacing the outcomes with missing values at $\approx 80\%$ of the spatial locations chosen uniformly at random and independently across the two margins of the outcome. In order to replicate the occurrence of regions with more sparsely observed outcomes, we replace outcomes with missing values at $\approx 99\%$ of spatial locations inside small circular areas whose center is chosen uniformly at random in $[0,1]^2$. 
Figure \ref{figure:synthetic_example} shows one of the resulting 900 data sets which we use to evaluate the performance of \modelnames. We consider multivariate \modelnames\ with $\delta = M$ and $\delta=1$ and compare them with multivariate cubic meshed GPs \citep[Q-MGPs;][]{meshedgp}, integrated nested Laplace approximations \citep[INLA;][]{inla} implemented via \texttt{R-INLA} using a $15\times 15$ grid, a low-rank multivariate GP method (labeled \textsc{lowrank}) on 25 knots obtained via \modelnames\ by setting $M=1$ with no domain partitioning, and an independent partitioning GP method (labeled \textsc{ind-part}) implemented by setting $M=1$ and partitioning the domain into 25 regions. Refer e.g. to \cite{Heaton2019} for an overview of low-rank and independent partitioning methods. We also include results from a non-spatial regression using Bayesian additive regression trees \citep[BART;][]{bart}. Each method was setup to target a compute time of 15 seconds for each data set. The total runtime for the 900 data sets thus amounted to about 26 hours. 

Tables \ref{table:synthetic_results:predictions} and \ref{table:synthetic_results:estimation} summarise the results across all 900 data sets. All Bayesian methods based on latent GPs exhibit very good coverage; in these simulated scenarios, \modelnames\ exhibit comparatively lower out-of-sample prediction errors. We highlight that the construction of DAG-based Bayesian methods for spatial regression depends on the underlying covariance function or kernel; comparisons of different covariance specifications across different methods is beyond the scope of this article.
Additional implementation details and figures can be found in Appendix \ref{appendix:synth_comparison}. 

\begin{table}
\centering
\resizebox{.60\columnwidth}{!}{%
\begin{tabular}{|l|rrr|}
  \hline
  Model & Cov. (95\%) & \textsc{RMSE}$(\by)$ & \textsc{MAE}$(\by)$ \\ 
  \hline
 \modelnames\ $\delta=M$  & 96.14 & \textbf{0.7171} & \textbf{0.5144} \\ 
 \textsc{inla} & 92.83 & 0.7521 & 0.5742 \\
  \textsc{q-mgp} & \textbf{95.87} & 0.7529 & 0.5410 \\ 
  \modelnames\ $\delta=1$ & 96.31 & 0.8027 & 0.5737 \\ 
  \textsc{bart} & 92.61 & 0.8820 & 0.6895 \\
  \textsc{lowrank} & 96.31 & 1.0230 & 0.7676 \\
  \textsc{ind-part} & 95.97 & 1.0796 & 0.8198\\
   \hline
\end{tabular}
}
\caption{Prediction performance on multivariate synthetic data: average coverage of 95\% prediction intervals, root mean square error (\textsc{RMSE}), and mean absolute error in prediction (\textsc{MAE}), over 900 data sets, sorted by lowest \textsc{RMSE}} \label{table:synthetic_results:predictions}
\end{table}

\begin{table}[ht]
\centering
\resizebox{.50\columnwidth}{!}{%
\begin{tabular}{|l|rr|}
  \hline
  Model & \textsc{RMSE}$(\by)$ & \textsc{MAE}$(\by)$ \\ 
  \hline
 \modelnames\ $\delta=M$   & \textbf{2.2882} & 1.6873\\ 
\textsc{lowrank}  & 2.6953 & \textbf{1.4895} \\
  \textsc{q-mgp}  & 2.8944 & 1.9630 \\ 
  \textsc{ind-part} &  2.9604 & 1.6079 \\
  \modelnames\ $\delta=1$ &  3.5883 & 2.3322 \\ 
   \hline
\end{tabular}
}
\caption{RMSE and MAE in the estimation of $\btheta$ from covariance function (\ref{eq:apanasovich_genton_covariance2}), averaged over 900 data sets.} \label{table:synthetic_results:estimation}
\end{table}

\subsection{Climate data: MODIS-TERRA and GHCN} \label{section:applications:modisnoaa}
Climate data are collected from multiple sources in large quantities; when originating from satellites and remote sensing, they are typically collected at high spatial and relatively low temporal resolution. Atmospheric and land-surface products are obtained via post-processing of satellite imaging, and their quality is negatively impacted by cloud cover and other atmospheric disturbances. On the other hand, data from a relatively small number of land-based stations is of low spatial but high temporal resolution. An advantage of land-based stations is that they measure phenomena related to atmospheric conditions which cannot be easily measured from satellites (e.g. precipitation data, depth of snow cover). 

We consider the joint analysis of five spatial outcomes collected from two sources. First, we consider Moderate Resolution Imaging Spectroradiometer (MODIS) data from the Terra satellite which is part of the NASA's Earth Observing System. Specifically, data product \texttt{MOD11C3} v. 6 provides monthly Land Surface Temperature (LST) values in a 0.05 degree latitude/longitude grid (the Climate Modeling Grid or CMG). The monthly data sets cover the whole globe from 2000-02-01 and consist of daytime and nighttime LSTs, quality control assessments, in addition to emissivities and clear-sky observations. 
The second source of data is the Global Historical Climatology Network (GHCN) database which includes climate summaries from land surface stations across the globe subjected to common quality assurance reviews. Data are published by the National Centers of Environmental Information (NCEI) of the National Oceanic and Atmospheric Administration (NOAA) at several different temporal resolutions; daily products report five core elements (precipitation, snowfall, snow depth, maximum and minimum temperature) in addition to several other measurements.

We build our data set for analysis by focusing on the continental United States in October, 2018. The MODIS data correspond to 359,822 spatial locations. Of these, 250,874 are collected at the maximum reported quality; we consider all remaining 108,948 spatial locations as missing, and extract (1) daytime LST (\texttt{LST\_Day\_CMG}), (2) nighttime LST (\texttt{LST\_Night\_CMG}), (3) number of days with clear skies (\texttt{Clear\_sky\_days}), (4) number of nights with clear skies (\texttt{Clear\_sky\_nights}). From the GHCN database we use daily data to obtain monthly averages for precipitation (\texttt{PRCP}), which is available at 24,066 spatial locations corresponding to U.S. weather stations; we log-transform \texttt{PRCP}. The two data sources do not share measurement locations as there is no overlap between measurement locations in MODIS and GHCN, with the latter data being collected more sparsely---this is a scenario of complete spatial misalignment. For this reason we build \modelnames\ favoring placement of GHCN locations at root nodes following Proposition \ref{prop:prop1}. Further implementation details are outlined at Appendix \ref{appendix:implementation}. 

\begin{figure}
    \centering
    \includegraphics[width=.85\textwidth]{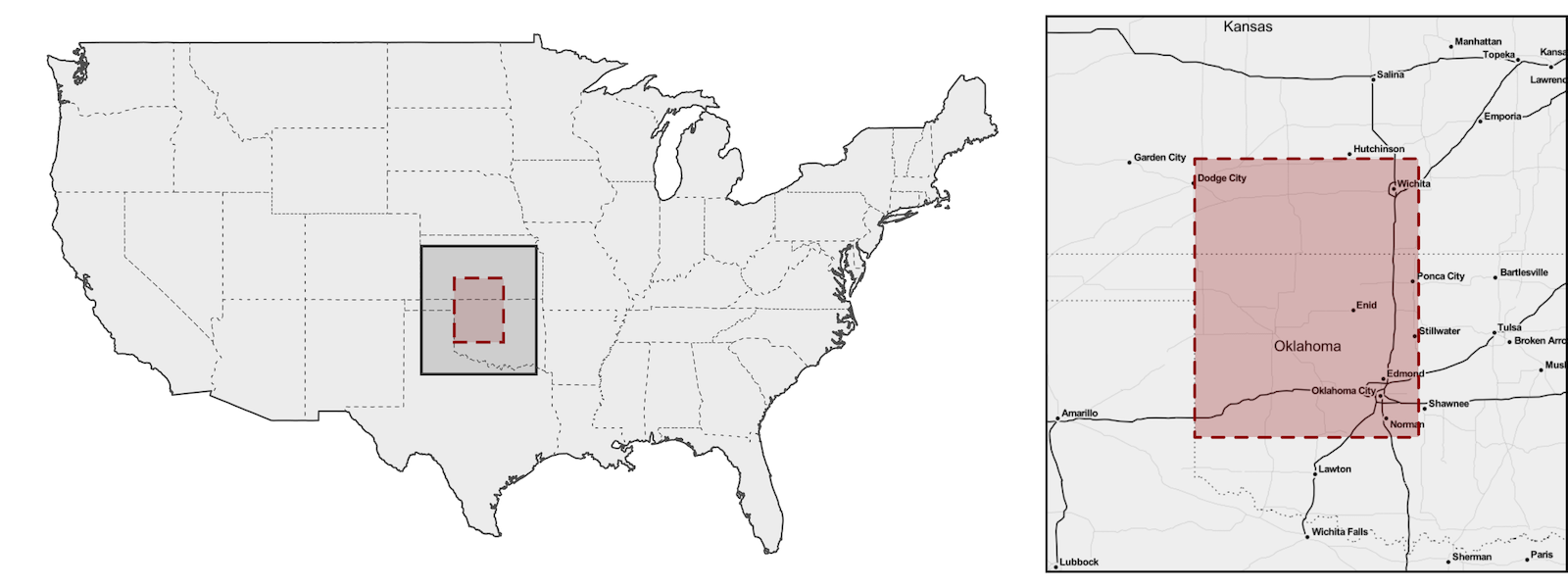}
    \caption{Prediction area}
    \label{fig:usamap}
\end{figure}

From the resulting data set of size $n=$1,027,562 we remove all observations in a large $3\times 3$ degree area in the central U.S. (from -100W to -97W and from 35N to 38N, i.e. the red area of Figure \ref{fig:usamap}) to build a test set on which we calculate coverage, MAE and RMSE of the predictions. We implement \modelnames\ on the covariance function (\ref{eq:apanasovich_genton_covariance2}).

\begin{figure}
\centering
\includegraphics[width=.95\textwidth]{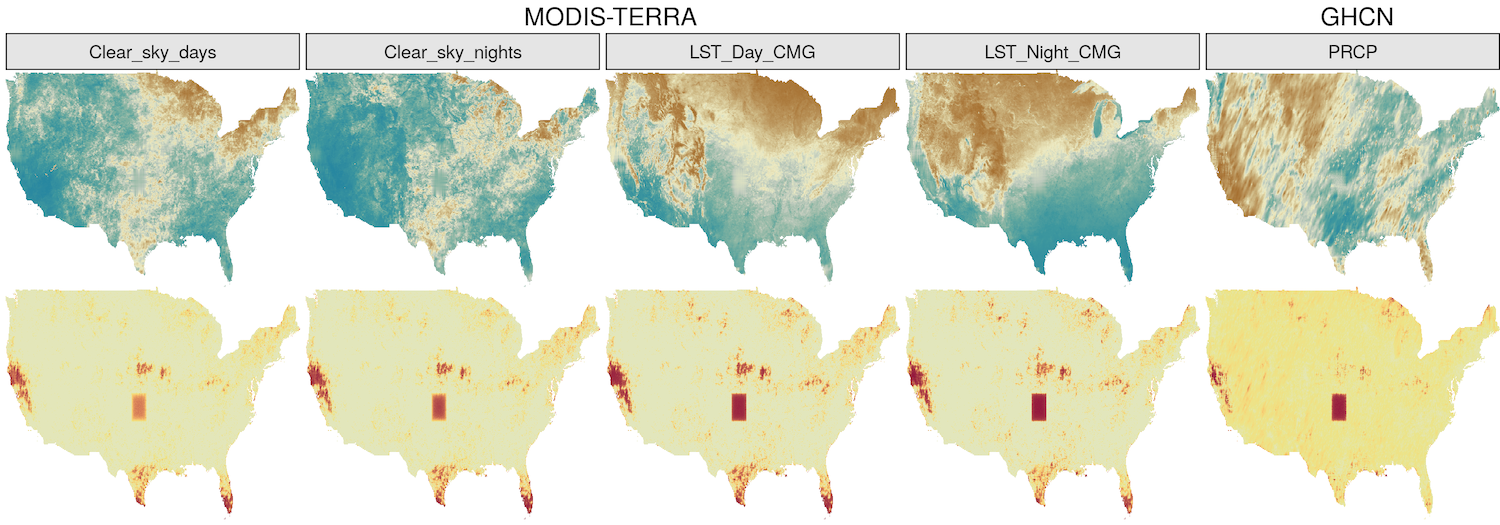}
\caption{Predicted values of the outcomes at all locations (top row) and associated 95\% uncertainty (bottom row), with darker spots corresponding to wider credible intervals.}
\label{fig:results_plot}
\end{figure}

\begin{figure}
    \centering
\includegraphics[width=.95\textwidth]{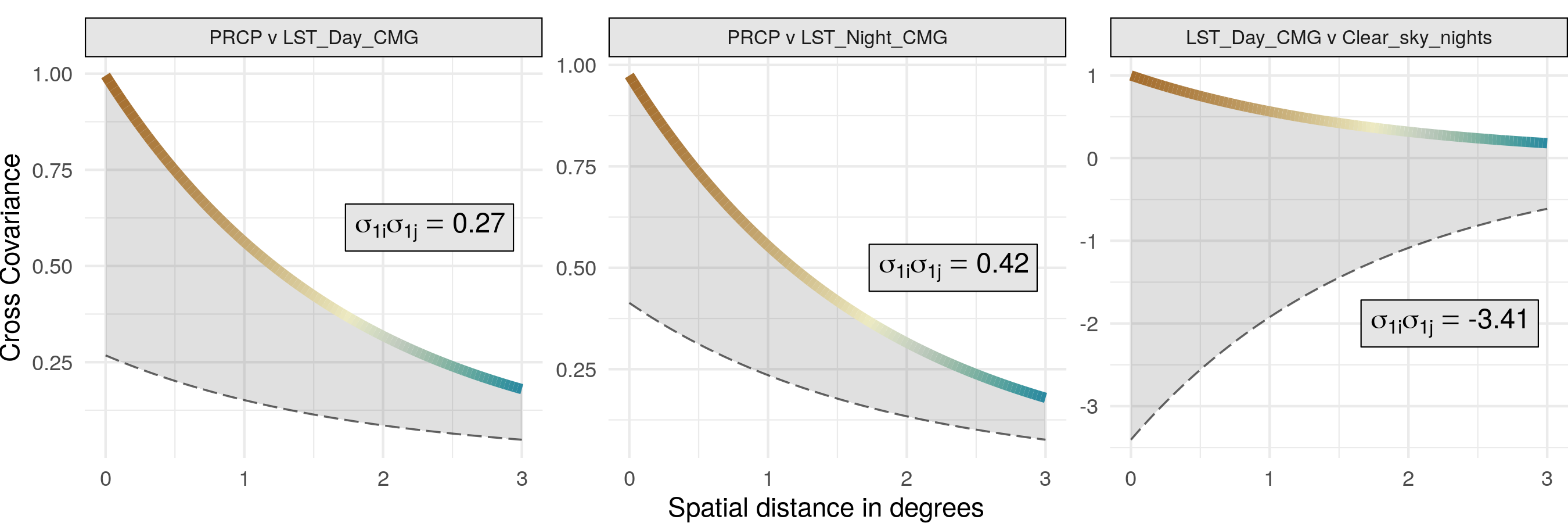}
    \caption{Given the latent dimensions $\delta_{ij}$, the color-coded lines represent $C(\bh, \delta_{ij})$ whereas $\Cov_{ij}(\bh) = \sigma_{1i} \sigma_{1j} C(\bh, \delta_{ij})$ is shown as a dashed grey line.}
    \label{fig:covariance_plot}
\end{figure}

Figure \ref{fig:results_plot} maps the predictions at all locations and the corresponding posterior uncertainties. Comparisons with other methods are difficult due to data size and complete misalignment. We implemented a tessellated MGP with the same covariance function and targeting similar computing time (predictive performance is reported in Appendix \ref{appendix:modisnoaa}); \modelnames\ displayed lower prediction errors and better coverage as seen in Table \ref{tab:predict_perf}.
We report posterior summaries of $\btheta$ in Appendix \ref{appendix:modisnoaa}. Opposite signs of $\sigma_{i1}$ and $\sigma_{j1}$ for pairs of variables $i,j \in \{1, \dots, q\}$ imply a negative relationship; however, the degree of spatial decay of these correlations is different for each pair as prescribed by the latent distances in the domain of variables $\delta_{ij}$. Figure \ref{fig:covariance_plot} depicts the resulting cross-covariance function for three pairs of variables. 

\begin{table}
\begin{tabular}{c}
\resizebox{16cm}{!}{
\begin{tabular}{|r|ccccc|}
\hline
Measure & \footnotesize \texttt{Clear\_sky\_days} & \footnotesize\texttt{Clear\_sky\_nights} & \footnotesize\texttt{LST\_Day\_CMG} &\footnotesize \texttt{LST\_Night\_CMG} & \texttt{PRCP} \\ 
\hline
95\% Coverage & 0.9798 & 0.9894 & 1.0000 & 0.9993 & 0.9717 \\ 
MAE & 1.2824 & 1.3029 & 0.9686 & 0.8440 & 0.3517 \\ 
RMSE & 1.6114 & 1.6214 & 1.2547 & 1.0764 & 0.5168 \\ 
\hline
\end{tabular}}
\\
\resizebox{16cm}{!}{
\begin{tabular}{|c|c|c|c|}
\hline
    $n=$1,014,017 &  Total iterations: 30,000 & Total time: 16.14h &
    Average time/iteration: 1.9s \\
    \hline
\end{tabular}}
\end{tabular}
\caption{Prediction results of \modelnames\ over the $3 \times 3$ degree area shown in Figure \ref{fig:usamap}}\label{tab:predict_perf}
\end{table}

\section{Discussion} \label{section:discussion}
In this article, we introduced \modelnames\ for Bayesian spatial multivariate regression modeling and provided algorithms for scalable estimation and prediction. \modelnames\ add significantly to the  class of methods for regression in spatially-dependent data settings.
We have demonstrated that \modelnames\ maintain accurate characterization of spatial dependence and scalability even in challenging settings involving multivariate data that are spatially misaligned.  Such complexities create problems for competing 
 approaches, including recent DAG-based approaches ranging from NNGPs to MGPs.

One potential concern is the need for users to choose a tree, and in particular specify 
 the number of locations associated to each node and the multivariate composition of locations in each node.  Although one can potentially estimate the tree structure based on the data, this would eliminate much of the computational speedup.  
We have provided theoretical guidance based on KL divergence from the full GP and computational cost associated to different tree structures.  This and our computational experiments lead to practical guidelines that can be used routinely in tree building.  Choosing a tree is simpler than the common task of choosing a neural network architecture in deep learning, and provides a useful degree of user-input to refine and improve upon an approach.  

We have focused on sampling algorithms for the latent effects because they provide a general blueprint which may be used for posterior computations in non-Gaussian outcome models; efficient algorithms for non-Gaussian big geostatistical data sets are currently lacking and are the focus of ongoing research.
Including time as a dimension is challenging and care must be taken when building a sparse DAG to avoid unreasonable assumptions on temporal dependence. For these reasons, future research may be devoted to building sparse DAG methods combining the advantages of treed structures with e.g. Markov-type assumptions of conditional independence.

\subsection*{Acknowledgements} This project has received funding from the European Research Council (ERC) under the European Union’s Horizon 2020 research and innovation programme (grant agreement No 856506).  This project was also partially funded by grant R01ES028804 of the United States National Institutes of Health (NIH).

\newpage

\vspace{2cm} 
\begin{center}
	{\Huge \textsc{\textbf{Appendix}}} 
\end{center}

\appendix

\section{Kolmogorov consistency conditions for \modelnames\ } \label{appx:kolmogorov}
We adapt results from \cite{nngp} and \cite{meshedgp}.
Let $w(\bs), \bs \in \mathcal{D}^*$ be the univariate representation in the augmented domain of the multivariate base process  $\{ \bw(\bl), \bl \in \calD \subset \Re^d \}$. Fix the reference set $\calS \subset \calD^*$ and let $\calL = \{ \bl_1, \dots, \bl_n \} \subset \calD^*$ and $\calU = \calL \setminus \calS$. Then 
\begin{align*}
\int \tp(\bw_{\calL}) \prod_{\bl_i \in \calL} dw(\bl_i) &=  \int \int \tilde{p}(\bw_{\calU} \mid \bw_{\calS}) \tilde{p}(\bw_{\calS}) \prod_{ \bs_i \in \calS \setminus \mathcal{L}} dw(\bs_i) \prod_{\bl_i \in \calL} dw(\bl_i)\\
&= \int \tp(\bw_{\calS}) \left( \int \tp(\bw_{\calU} \mid \bw_{\calS}) \prod_{\bl_i \in \calU} dw(\bl_i) \right) \prod_{\bl_i\in \calS} dw(\bl_i) = 1,
\end{align*}
hence $\tp(\bw_{\calL})$ is a proper joint density. To verify the Kolmogorov consistency conditions, take the permutation $\calL_{\pi} = \{ \bl_{\pi(1)}, \dots,  \bl_{\pi(n)} \}$ and call $\calU_{\pi} = \calL_{\pi} \setminus \calS$. Clearly $\calU_{\pi} = \calL_{\pi} \setminus \calS = \calL \setminus \calS = \calU$ and similarly $\calS \setminus \calL_{\pi} = \calS \setminus \calL$ so that 
\begin{align*}
\tilde{p}(\bw_{\calL_\pi}) &= \int \tilde{p}(\bw_{\calU_\pi} \mid \bw_{\calS}) \tilde{p}(\bw_{\calS}) \prod_{ \bs_i \in \calS \setminus \calL_\pi} dw(\bs_i) \\
&= \int \tilde{p}(\bw_{\calU} \mid \bw_{\calS}) \tilde{p}(\bw_{\calS}) \prod_{ \bs_i \in \calS \setminus \calL} dw(\bs_i) = \tp(\bw_\calL)
\end{align*}
implying
\begin{align*}
\tilde{p}(\bw(\bl_1), \dots, \bw(\bl_n)) &= \tilde{p}(\bw(\bl_{\pi(1)}), \dots, \bw(\bl_{\pi(n)})).
\end{align*}
Next, take a new location location $\bl_0 \in \calD^*$.  Call $\calL_1 = \calL \cup \{ \bl_0 \} $. We want to show that $\int \tp(\bw_{\calL_1}) d w(\bl_0) = \tp(\bw_{\calL}) $. If $\bl_0 \in \calS$ then $\calL_1 \setminus \calS = \calL\setminus \calS = \calU$ and hence 
\begin{align*}
\int \tp(\bw_{\calL_1}) dw(\bl_0) &= \int \left(  \tp( \bw_{\calL_1 \setminus \calS} \mid \bw_{\calS} ) \tp(\bw_{\calS})\prod_{\bs_i \in \calS\setminus \calL_1} dw(\bs_i) \right) dw(\bl_0) \\
&= \int  \tp( \bw_{\calU} \mid \bw_{\calS} ) \tp(\bw_{\calS}) \prod_{\bs_i \in \calS\setminus \calL} d w(\bs_i) = \tp(\bw_{\calL}).
\end{align*}
If $\bl_0 \notin \calS$ we have
\begin{align*}
\int \tilde{p}(\bw_{\calL_1}) d w(\bl_0) &= \int \left( \int \tilde{p}(\bw_{\calL_1 \setminus \calS} \mid \bw_{\calS}) \tilde{p}(\bw_{\calS}) \prod_{ \bs_i \in \calS \setminus \calL_1} dw(\bs_i) \right) d w(\bl_0)\\
&= \int \left( \int \tilde{p}(\bw_{\calL \setminus \calS \cup \{ \bl_0 \}} \mid \bw_{\calS}) \tilde{p}(\bw_{\calS}) \prod_{ \bs_i \in \calS \setminus \calL} d w(\bs_i) \right) d w(\bl_0) \\
&= \int \left( \int \tilde{p}(\bw_{\{ \bl_0 \}} \mid \bw_{\calL \setminus \calS}, \bw_{\calS}) \tilde{p}(\bw_{\calL \setminus \calS} \mid \bw_{\calS}) \tilde{p}(\bw_{\calS}) \prod_{ \bs_i \in \calS \setminus \calL} dw(\bs_i) \right) d w(\bl_0)\\
&= \int \tilde{p}(\bw_{\calL \setminus \calS} \mid \bw_{\calS}) \tilde{p}(\bw_{\calS}) \prod_{ \bs_i \in \calS \setminus \calL} dw(\bs_i)\int \tilde{p}(\bw_{\{ \bl_0 \}} \mid \bw_{\calS}) dw(\bl_0) \\
&=\int \tilde{p}(\bw_{\calL \setminus \calS} \mid \bw_{\calS}) \tilde{p}(\bw_{\calS}) \prod_{ \bs_i \in \calS \setminus \calL} d w(\bs_i)\\
&=\tilde{p}(\bw_{\calL}).
\end{align*}

\section{Properties of Gaussian \modelnames\ }\label{appendix:gp_properties}
Consider the treed graph $\calG$ of a \modelname. In this section, we make no distinction between reference and non-reference nodes, and instead label $\bV_i = \bA_i$ for $i=0, \dots, M-1$ and $\bV_M = \bB$ so that $\bV = \{ \bA, \bB \} = \{ \bV_0, \dots, \bV_{M-1}, \bV_M \}$ and the $\bV_M$ are the leaf nodes. Each $\bw_i$ is $n_i \times 1$ and corresponds to $\bv_i \in \bV_r$ for some $r=0, \dots, M$ so that $\pa{\bv_i} = \{ \bv_{j_1}, \dots, \bv_{j_r} \}$ for some sequence $\{ j_1, \dots, j_r \}$, 
and $\eta^{-1}(\pa{\bv_i}) = \{ S_{j_1}, \dots, S_{j_r} \}$. Denote the $h$-th parent of $\bv_i$ as $\pa{\bv_i}(h)$. 

\subsection{Building the precision matrix}\label{section:build_precision}
We can represent each conditional density $N( \bw_i \mid \bH_{i} \bwpa{i}, \bR_{i})$ as a linear regression on $\bw_i$:
\begin{equation}\label{section:general:eq:cond_reg}
\bw_0 = \bolds{\omega}_0 \sim N(\bolds{0},\bR_0), \quad \bw_i = \sum_{\{j : \bv_j\in \pa{\bv_i}\}} \bh_{ij}\bw_j + \bolds{\omega}_i\;,\;\; i=1,2,\ldots,M,
\end{equation}
where each $\bh_{ij}$ is an $n_i \times n_j$ coefficient matrix representing the regression of $\bw_i$ given $\bw_{[i]}$, $\bolds{\omega}_i \stackrel{ind}{\sim} N(\bolds{0}, \bR_i)$ for $i=0, 1, \ldots, M$, and each $\bR_i$ is an $n_i\times n_i$ residual covariance matrix. We set $\bh_{ii}=\bO$ and $\bh_{ij}=\bO$, where $\bO$ is the matrix of zeros, whenever $j \notin \{ j_1, \dots, j_r \}$. Using this representation, we have $\bH_{i} = \left[\bh_{i,j_1},\bh_{i,j_2},\ldots,\bh_{i,j_{r}}\right]$, which is an $n_i \times J_i$ block matrix formed by stacking $\bh_{i,j_k}$ side by side for $k=1, \dots, r$. Since $\mbox{E}[\bw_i\given\bw_{[i]}] = \bH_{i}\bw_{[i]} = \Cov_{i,[i]}\Cov_{[i]}^{-1}\bw_{[i]}$, we obtain $\bH_{i} = \Cov_{i,[i]}\Cov_{[i]}^{-1}$. We also obtain $\bR_i = \mbox{var}\{\bw_i\given \bw_{[i]}\} = \Cov_{i,i} - \Cov_{i,[i]}\Cov_{[i]}^{-1}\Cov_{[i],i}$, hence all $\bH_i$'s, $\bh_{ij}$'s, and $\bR_i$'s can be computed from the base covariance function. 

In order to continue building the precision matrix, define the block matrix $\bbH = \{ \bh_{ij }\}$. We can write
\begin{equation} \label{equation:hij_matrices}
    \bh_{ij} = \begin{cases}
        \bO & \text{ if } \bv_j\notin \pa{\bv_i} \\
        (\Cov_{i, [i]} \Cov^{-1}_{[i]})(\cdot, h) = \bH_i(\cdot, h) & \text{ if } \bv_j = \bv_{j_h} \in \pa{\bv_i},
    \end{cases}
\end{equation}
where $(\cdot,h)$ refers to the $h$-th block column. More compactly using the indicator function $\mathbf{1}\{ \cdot  \}$ we have $\bh_{ij} = \mathbf{1}\{\exists h : \bv_j = \pa{\bv_i}(h)\} (\Cov_{i, [i]} \Cov^{-1}_{[i]})(\cdot, h)$.
If we stack \textit{all} the $\bh_{ik}$ horizontally for $k=0, \dots, M_S-1$, we obtain the $n_i \times n$ matrix $\bbH(i,\cdot)$, which is $i$-th block row of $\bbH$. 
Intuitively, $\bbH(i,\cdot)$ is a \textit{sparse} matrix with the coefficients linking the full $\bw$ to $\bw_i$, with zero blocks at locations whose corresponding node is $\bv_j \notin \pa{\bv_i}$. The $i$th block-row of $\bbH$ is of size $n_i \times n$ but only has $r$ non-empty sub-blocks, with sizes $n_i \times n_j$ for $j \in \{ j_1, \dots, j_r \}$, respectively. Instead, $\bH_i$ is a \textit{dense} matrix obtained by dropping all the zero-blocks from $\bbH(i,\cdot)$, and stores the coefficients linking $\bwpa{i}$ to $\bw_i$. The two are linked as $\bH_i \bwpa{i} = \bbH(i, \cdot) \bw$.

Since $\bw = \bbH\bw + \bolds{\omega}$, $\tilde{\Cov} = \mbox{var}(\bw) = (\bI - \bbH)^{-1}\bR(\bI - \bbH)^{-\top}$, where $\bR = \bdiag\{ \bR_i \}$ and $\bI-\bbH$ is block lower-triangular with unit diagonal, hence non-singular. We find the precision matrix as $\tilde{\Cov}^{-1} = (\bI - \bbH)^{\top}\bR^{-1}(\bI-\bbH)$.

\subsection{Properties of $\tilde{\Cov}^{-1}$}
When not ambiguous, we use the notation $\bX_{ij}$ to denote the $(i,j)$ block of $\bX$. An exception to this is the $(i,j)$ block of $\tilde{\Cov}^{-1}$ which we denote as $\tilde{\Cov}^{-1}(i,j)$. In \modelnames, $\tilde{\Cov}^{-1}(i,j)$ is nonzero if $i=j$ or if the corresponding nodes $\bv_i$ and $\bv_j$ are connected in the moral graph $\mathcal{G}^m$, which is an undirected graph based on $\calG$ in which an edge connects all nodes that share a child. This means that either (1) $\bv_i \in \pa{\bv_j}$ or vice-versa, or (2) there exists $\ba^*$ such that $\{ \bv_i, \bv_j \} \subset \pa{\ba^*}$. In \modelnames, $\calG^m = \calG$. In fact, suppose there is a node $\ba^* \in \bv_{r^*}$ such that $\ba^* \in \ch{\bv_j} \cap \ch{\bv_k}$, where $\bv_j \in \bv_{r_j}$ and $\bv_k \in \bv_{r_k}$. By definition of $\calG$ there exists a sequence $\{i_1, \dots, i_{r^*} \}$ such that $\pa{\ba^*} = \{ \bv_{i_1}, \dots, \bv_{i_{r^*}} \} \supset \{ \bv_j, \bv_k \}$, and furthermore $\pa{\bv_{i_h}} = \{\bv_{i_1}, \dots, \bv_{i_{h-1}} \}$ for $h \leq r^*$. This implies that if $j=k$ then $\bv_j=\bv_k$, whereas if $j>k$ then $\bv_k \in \pa{\bv_j}$, meaning that no additional edge is necessary to build $\calG^m$.

\subsubsection{Explicit derivation of $\tilde{\Cov}^{-1}(i,j)$}
Denote $\bR^{-1} = \bR^{-\frac{1}{2}} \bR^{-\frac{\top}{2}}$, $\bU = (\bI - \bbH)^\top \bR^{-\frac{1}{2}}$, and define the ``common descendants'' as $\text{cd}(\bv_i, \bv_j) = (\{ \bv_i \} \cup \ch{\bv_i}) \cap (\{ \bv_j \} \cup \ch{\bv_j})$. Then consider $\ba_i \in \bA, \bv_j \in \bV$ such that $\ba_i \in \pa{\bv_j}$ and denote as $\bH_{i\to j}$ the matrix obtained by subsetting $\bH_j$ to columns corresponding to $\ba_i$ and note that $\bH_{i\to j} = \bbH_{ji}$.
The $(i,j)$ block of $\bU$ is then
\begin{equation*}
    \bU_{ij} = \begin{cases}
        \bO_{n_i \times n_j} & \text{ if } \bv_j\notin \ch{\bv_i} \\
        \bI_{n_i \times n_i} & \text{ if } i = j \\
        (\bI_{ji} - \bH_{i\to j})^\top \bR_j^{-\frac{1}{2}} & \text{ if } \bv_j \in \ch{\bv_i}\\
        \quad\quad = (\bI_{ji} - \bbH_{ji})^\top \bR_j^{-\frac{1}{2}}
    \end{cases}
\end{equation*}
Then $\tilde{\Cov}^{-1}(i,j) = \sum_k \bU_{ik} \bU_{jk}^{\top}$ and, as in (\ref{equation:precision_blocks_main}), each block of the precision matrix is:
\begin{equation}\label{equation:precision_blocks}
\begin{aligned}
\tilde{\Cov}^{-1}(i, j) &= \sum\limits_{\bv_k \in \text{cd}(\bv_i, \bv_j)}
    (\bI_{ki} - \bH_{i\to k})^\top \bR_k^{-1} (\bI_{kj} - \bH_{j\to k})\\
    &= \sum\limits_{\bv_k \in \text{cd}(\bv_i, \bv_j)}
    (\bI_{ki} - \bbH_{ki})^\top \bR_k^{-1} (\bI_{kj} - \bbH_{kj})
\end{aligned}
\end{equation}
where $\text{cd}(\bv_i, \bv_j) = \emptyset$ implies $\tilde{\Cov}^{-1}(i,j) = \bO$ and $\bI_{ij}$ a zero matrix unless $i=j$ as it is the $(i,j)$ block of an identity matrix of dimension $n\times n$.

\subsubsection{Computation of large matrix inverses}
One important aspect in building $\tilde{\Cov}^{-1}$ is that it requires the computation of the inverse $\Cov^{-1}_{[i]}$ of dimension $J_i \times J_i$ for all nodes with parents, i.e. at $r>0$. Unlike models which achieve scalable computations by limiting the size of the parent set (e.g. NNGPs and their blocked variant, or tessellated MGPs), this inverse is increasingly costlier when $\delta>1$ for nodes at a higher-level of the tree as those nodes have more parents and hence larger sets of parent locations (the same conclusion holds for non-reference nodes). However, the treed structure in $\calG$ allows one to avoid computing the inverse in $O(J_i^3)$. In fact, suppose we have a symmetric, positive-definite block-matrix $\bA$ and we wish to compute its inverse. We write
\begin{align*}
    \bA = \begin{bmatrix} C & B \\
    B^\top & D\end{bmatrix} & \quad \bA^{-1} = \begin{bmatrix} C^{-1} + C^{-1}BS^{-1}B^\top C^{-1} & -C^{-1} B S^{-1} \\
    -S^{-1}B^\top C^{-1} & S^{-1}\end{bmatrix},
\end{align*}
where $S = C - BD^{-1}B^\top$ is the Schur complement of $D$ in $\bA$. If $C^{-1}$ was available, the only necessary inversion is that of $S$. In \modelnames\ with $\delta>1$,  suppose $\bv_i, \bv_j$ are two nodes such that $\pa{\bv_j} = \{ \bv_i \} \cup \pa{\bv_i}$ -- this arises for nodes $\bv_j \in \bV_r, r \geq M_{\delta}$. Regardless of whether $\bv_j$ is a reference node or not, $\eta^{-1}( \pa{\bv_j} ) = \{\calS_i, \calS_{[i]} \}$ and
\begin{align*}
    \Cov_{[j]} = \begin{bmatrix} \Cov_{[i]} & \Cov_{[i], i} \\
    \Cov_{i, [i]} & \Cov_{i} \end{bmatrix}, & \quad \Cov_{[j]}^{-1} = \begin{bmatrix} \Cov_{[i]}^{-1} + \Cov_{[i]}^{-1}\Cov_{[i], i}S^{-1}\Cov_{i, [i]} \Cov_{[i]}^{-1} & -\Cov_{[i]}^{-1} \Cov_{[i], i} S^{-1} \\
    -S^{-1}\Cov_{i, [i]} \Cov_{[i]}^{-1} & S^{-1}\end{bmatrix},
\end{align*}
where the Schur complement of $\Cov_i$ is $S = \Cov_{i} - \Cov_{i, [i]} \Cov_{[i]}^{-1} \Cov_{[i], i} = \bR_i$. Noting that $\bH_i = \Cov_{i, [i]}\Cov^{-1}_{[i]}$ we write
\begin{align} \label{equation:nested_inverse}
    \Cov_{[j]}^{-1} = \begin{bmatrix} \Cov_{[i]}^{-1} + \bH_i^\top \bR_i^{-1}\bH_i & - \bH_i^\top \bR_i^{-1} \\
    -\bR_i^{-1}\bH_i  & \bR_i^{-1}\end{bmatrix}.
\end{align}

\subsubsection{Computing $(\tilde{\Cov}^{-1} + \bSigma)^{-1}$ and its determinant without sparse Cholesky} \label{appendix:precision_decomposition}
Bayesian estimation of regression models requiring the computation of $(\bZ \tilde{\Cov} \bZ^\top + \bD)^{-1}$ and its determinant use the Sherman-Morrison-Woodbury matrix identity to find $(\bZ \tilde{\Cov} \bZ^\top + \bD)^{-1} = \bD^{-1} - \bD^{-1} \bZ ( \tilde{\Cov}^{-1} + \bSigma )^{-1} \bZ^\top \bD^{-1}$, where $\bSigma = \bZ^\top \bD^{-1} \bZ$. A sparse Cholesky factorization of $\tilde{\Cov}^{-1} + \bSigma$ can be used as typically $\bSigma$ is diagonal or block-diagonal, thus maintaining the sparsity structure of $\tilde{\Cov}^{-1}$. Sparse Cholesky libraries \citep[e.g. \texttt{Cholmod},][]{cholmod}, which are embedded in software or high-level languages such as \textsc{Matlab}\texttrademark\ or the \texttt{Matrix} package for \texttt{R}, scale to large sparse matrices but are either too flexible or too restrictive in our use cases: (1) we know $\calG$ and its properties in advance; (2) \modelnames\ take advantage of block structures and grouped data. In fact, sparse matrix libraries typically are agnostic of $\calG$ and heuristically attempt to infer a sparse $\calG$ given its moralized counterpart. While this operation is typically performed once, \textit{a priori} knowledge of $\calG$ implies that reliance on such libraries is in principle unnecessary. 

We thus take advantage of the known structure in $\calG$ to derive direct algorithms for computing $(\tilde{\Cov}^{-1} + \bSigma)^{-1}$ and its determinant. In the discussion below we consider $\delta=M$, noting here that choosing $\delta=1$ simplifies the treatment as $\text{cd}(\bv_i, \bv_j) = \{ \bv_i \}$ if $\bv_i=\bv_j$, and it is empty otherwise. 
We now show how (\ref{equation:precision_blocks}) leads to Algorithm \ref{algorithm:precision_decomposition} for the decomposition of \textit{any} precision matrix $\bLambda$ which conforms to $\calG$ -- i.e. it has the same block-sparse structure as a precision matrix built as in Section \ref{section:build_precision}. Suppose from $\bLambda$ we seek a block lower-triangular matrix $\bL$ and a block diagonal $\bD$ such that 
\[\bLambda_{ij} = \sum\limits_{\bv_k \in \text{cd}(\bv_i, \bv_j)}(\bI_{ki} - \bL_{ki})^\top \bD_k (\bI_{kj} - \bL_{kj}).\] 
Start with $\bv_i, \bv_j$ taken from the leaf nodes, i.e. $\bv_i, \bv_j \in \bV_{M}$. Then $\text{cd}(\bv_i, \bv_j) = \emptyset$ and we set $\bL_{ij} = \bL(j,i)^\top = \bO = \bLambda_{ij}$. If $i=j$ then $\text{cd}(\bv_i, \bv_i) = \{ \bv_i \}$ and
\begin{align*}
   \sum_{\bv_k \in \text{cd}(\bv_i, \bv_i)} &(\bI_{ki} - \bL_{ki})^\top \bD_k (\bI_{ki} - \bL_{ki}) = (\bI_{ii} - \bL_{ii})^\top \bD_i (\bI_{ii} - \bL(i, i)) \\
   &= \bD_i - \bD_i \bL_{ii} - \bL_{ii}^{\top} \bD_i + \bL_{ii}^{\top} \bD_i \bL_{ii};
\end{align*}
we then set $\bL_{ii} = \bO$ and get the $i$-th block of $\bD_i$ simply setting $\bD_i = \bLambda_{ii}$. Proceeding downwards along $\calG$, if $\bv_j \in \bV_{M-1} \cap \pa{\bv_i}$ we have $\bLambda_{ij} = \bD_i(\bI_{ij} - \bL_{ij}) = - \bD_i \bL_{ij} $ and thus set $\bL_{ij} = -\bD_i^{-1} \bLambda_{ij}$. We then note that $\text{cd}(\bv_j, \bv_j) = \{ \bv_j, \bv_i \}$ and obtain $\bLambda_{jj} = \bD_j + \bL_{ij}^\top \bD_i \bL_{ij}$ where $\bL_{ij}$ and $\bD_i$ have been fixed at the previous step; this results in $\bD_j = \bLambda_{jj} - \bL_{ij}^\top \bD_i \bL_{ij}$. 

Then, the $s$-th (of $M$) step takes $\bv_j \in \bV_{M-s} \cap \pa{\bv_i}$ and $\bv_i \in \bV_{M-s+1}$, implying $\text{cd}(\bv_i, \bv_j) = \{ \bv_i \} \cup \ch{\bv_i}$. Noting that $\bF^* = \sum_{\bv_k \in \ch{\bv_i}}  (\bI_{ki} - \bL_{ki})^\top \bD_k (\bI_{kj} - \bL_{kj})$ has been fixed at previous steps since each $\bv_k$ is at level $M-s+2$, we split the sum in (\ref{equation:precision_blocks}) and get
\begin{align*}
\bLambda_{ij} &- \bF^* = \bD_i (\bI_{ij} - \bL_{ij}) = -\bD_i \bL_{ij} ,
\end{align*}
where $\bD_i$ has been fixed at step $s-1$, obtaining $\bL_{ij} = -\bD_i^{-1} (\bLambda_{ij} - \bF^*)$; $\bD_j$ can be found using the same logic. Proceeding until $M-s = 0$ from the leaves of $\calG$ to the root, we ultimately fill each non-empty block in $\bL$ and $\bD$ resulting in $\bLambda = (\bI - \bL)^{\top} \bD (\bI - \bL)$. Algorithm \ref{algorithm:precision_decomposition} unifies these steps to obtain the block decomposition of any sparse precision matrix $\bLambda$ conforming to $\calG$ resulting in $\bLambda = (\bI - \bL)^\top \bD (\bI - \bL)$, where $\bL$ is block lower triangular and $\bD$ is block diagonal. 
This is akin to a block-LDL decomposition of $\bLambda$ indexed on nodes of $\calG$.
Algorithm \ref{algorithm:precision_decomp_invtri} complements this decomposition by providing a $\calG$-specific block version of forward substitution for computing $(\bI - \bL)^{-1}$ with $\bL$ as above. 

In practice, a block matrix with $K^2$ blocks can be represented as a $K^2$ array with rows and columns indexed by nodes in $\calG$ and matrix elements which may be zero-dimensional whenever corresponding to blocks of zeros. The specification of all algorithms in block notation allows us to never deal with large (sparse) matrices in practice but only with small block matrices indexed by nodes in $\calG$, bypassing the need for external sparse matrix libraries. Specifically we use the above algorithms to compute $\bLambda^{-1} = (\tilde{\Cov}^{-1} + \bSigma)^{-1}$ and its determinant: $\bLambda^{-1} = (\bI - \bL)^{-1} \bD^{-1} (\bI-\bL)^{-\top}$
and $|\bLambda^{-1}| = \prod_{i=1}^{M_S} 1/|\tilde{\bD}_i|$. We have not distinguished non-reference and reference nodes in this discussion. In cases in which the non-reference set is large, we note that the conditional independence of all non-reference locations, given their parents, results in $\tilde{\Cov}^{-1}(i,i)$ being diagonal for all $\bl \in \calU$ (i.e. $\eta(\bl) = \bv_i \in \bB$). This portion of the precision matrix can just be stored as a column vector.

\begin{algorithm}
\SetKwInOut{Input}{Input}
\Input{$\bLambda$ $n\times n$ precision matrix conforming to $\calG$}
 Initialize $\bL=\bO_{n\times n}, \bD=\bO_{n\times n}$\;
 \For(\tcp*[f]{top down from last level}){$r \in \{M, \dots, 0\}$}{
    \For(\tcp*[f]{[parallel for]}){$j : \{ \bv_j \in \bV_r\}$}{
        $\bD_{jj} = \bLambda_{jj}$\;
        \For{$p : \{ \bv_{p} \in \pa{\bv_j} \}  $}{
            $\bL_{jp} = - \bD^{-1}_{jj} \bLambda_{jp}$\;
            \For{$g : \{ \bv_{g} \in \pa{\bv_j} \} $ }{
            $\bLambda_{pg} = \bLambda_{pg} - \bLambda_{jp}^\top \bL_{jg}$\;
            $\bLambda_{gp} = \bLambda_{pg}^\top$\;
            }
        }
    }
 }
 \textbf{Result:} Block-lower-triangular $\bL$ with $\bL_{ij} \neq \bO$ if $\bv_i \in \pa{\bv_j}$, and block-diagonal $\bD$ such that $(\bI - \bL)^\top \bD (\bI - \bL) = \bLambda$.
 \caption{Precision matrix decomposition given treed graph $\calG$ with $M$ levels.}\label{algorithm:precision_decomposition}
\end{algorithm}

\begin{algorithm}
\SetKwInOut{Input}{Input}
\Input{$\bGamma = \bI - \bL$ where $\bL$ is as in Algorithm \ref{algorithm:precision_decomposition}.}
Initialize $\bDelta_{ij} = \bO_{n_i, n_j}$ for all $i, j$ such that $\bv_j \in \pa{\bv_i}$\;
 \For(\tcp*[f]{bottom up from root of $\calG$}){$r \in \{0, \dots, M\}$}{
    \For(\tcp*[f]{[parallel for]}){$j : \{ \bv_j \in \bV_r\}$}{
        \For{$p : \{ \bv_{p} \in \tilde{\calP}_{0\to [j]} \}  $}{
            Set $\text{chain}(\bv_{p}, \bv_j) = \{\bv_{p}\} \cup \{ \tilde{\calP}_{0\to [j]} \cap \tilde{\calP}_{0\to [p]}\} $\;
            \For{$g : \{ \bv_{g} \in \text{\normalfont chain}(\bv_{p}, \bv_j) \} $ }{
                $\bDelta_{jp} = \bDelta_{jp} - \bGamma_{jg} \bDelta_{gp}$
            }
        }
    }
 }
 \textbf{Result:} $\bDelta = \bGamma^{-1}$.
 \caption{Calculating the inverse of $\bI - \bL$ with $\bL$ output from Algorithm \ref{algorithm:precision_decomposition}.}\label{algorithm:precision_decomp_invtri}
\end{algorithm}

\subsubsection{Sparsity of $\tilde{\Cov}^{-1}$}\label{appendix:sparsity}
We calculate the sparsity in the precision matrix; considering an enumeration of nodes by level in $\calG$, denote $n_{ij} = |\eta^{-1}(\bv_{ij})|$, $m_j = |\bV_j|$, and $J_{ij} = | \eta^{-1}(\pa{\bv_{ij}})|$, and noting that by symmetry $(\tilde{\Cov}^{-1}(i,j))^{\top} = \tilde{\Cov}^{-1}(j,i)$, the number of nonzero elements of $\tilde{\Cov}^{-1}$ is 
\[ \text{nnz}(\tilde{\Cov}^{-1}) = \sum_{j=0}^{M} \sum_{i=1}^{m_j} \left( 2n_{ij} J_{ij} + n_{ij}^2\bolds{1}\{j<M\} + n_{ij} \bolds{1}\{j=M\} \right),\]
where $n_{ij} \bolds{1}\{ j=M \}$ refers to the diagonal elements of the precision matrix at non-reference locations.

\subsection{Properties of \modelnames\ with $\delta=M$}
We outline recursive properties of $\tilde{\Cov}$ induced by $\calG$ when $\delta=M$. In the case $1< \delta< M$, these properties hold for nodes at or above level $M_{\delta}$, using $\bA_{M_{\delta}}$ as root. We focus on paths in $\calG$. These can be represented as sequences of nodes $\{ \bv_{i_1}, \dots, \bv_{i_r} \}$ such that $\{ \bv_{i_j}, \dots, \bv_{i_k}\} \subset \pa{\bv_{i_{k+1}}}$ for $1<j<k<r$. Take two successive elements of such a sequence, i.e. $\bv_i, \bv_j$ such that $\bv_i \rightarrow \bv_j$ in $\calG$. Consider $\mbox{E}[\bw_j\given\bw_{[j]}] = \bH_{j}\bw_{[j]} = \Cov_{j,[j]}\Cov_{[j]}^{-1}\bw_{[j]}$ and $\bR_j = \mbox{var}\{\bw_j\given \bw_{[j]}\} = \Cov_{j,j} - \Cov_{j,[j]}\Cov_{[j]}^{-1}\Cov_{[j],j}$. By (\ref{equation:nested_inverse}) we can write
\begin{align*}
    \bH_j & \bwpa{j} = \begin{bmatrix} \Cov_{j, [i]} & \Cov_{j, i} \end{bmatrix}  
    \begin{bmatrix} \Cov_{[i]}^{-1} + \bH_i^\top  \bR_i^{-1}\bH_i & - \bH_i^\top \bR_i^{-1} \\
    -\bR_i^{-1}\bH_i & \bR_i^{-1}\end{bmatrix} \begin{bmatrix} \bwpa{i} \\ \bw_{i}  \end{bmatrix} \\
    = &
    \begin{bmatrix} \Cov_{j,[i]} & \Cov_{j,i} - \Cov_{j,[i]}\Cov^{-1}_{[i]}\Cov_{[i],i} \end{bmatrix}
    \begin{bmatrix} \Cov^{-1}_{[i]} & \bO\\ \bO & (\Cov_{i,i} - \Cov_{i,[i]}\Cov^{-1}_{[i]}\Cov_{[i],i})^{-1} \end{bmatrix}
    \begin{bmatrix} \bwpa{i} \\ \bw_i - \Cov_{i,[i]} \Cov^{-1}_{[i]} \bw_{[i]}  \end{bmatrix} \\
    = &
    \begin{bmatrix} \Cov_{j,[i]}\Cov^{-1}_{[i]} & (\Cov_{j,i} - \Cov_{j,[i]}\Cov^{-1}_{[i]}\Cov_{[i],i})(\Cov_{i,i} - \Cov_{i,[i]}\Cov^{-1}_{[i]}\Cov_{[i],i})^{-1} \end{bmatrix}
    \begin{bmatrix} \bwpa{i} \\ \bw_i - \Cov_{i,[i]} \Cov^{-1}_{[i]} \bw_{[i]}  \end{bmatrix}. 
\end{align*}
Now define the covariance function $\bK_i(\bl, \bl') = \Cov_{\bl, \bl'} - \Cov_{\bl, [i]} \Cov^{-1}_{[i]} \Cov_{[i], \bl'}$; recalling that the reference set is $\calS = \cup_{i=0}^{M-1} \cup_{j=1}^{m_j} S_j$ we use a shorthand notation for these subsets: $\bK_i(S_h, S_k) = \bK_i(h,k)$. Also denote $\be_i = \bw_i - \Cov_{i,[i]}\Cov^{-1}_{[i]} \bwpa{i} $ for all $i$.
The above expression becomes
\begin{align*}
    \bH_j \bwpa{j} &= \begin{bmatrix} \Cov_{j,[i]}\Cov^{-1}_{[i]} & \bK_{i}(j,i) \bK_{i}^{-1}(i,i) \end{bmatrix}
    \begin{bmatrix} \bwpa{i} \\ \be_i  \end{bmatrix}\\
    &= \bH_i \bwpa{i} + \bK_{i}(j,i) \bK_{i}^{-1}(i,i) \be_i;
\end{align*}
we can use this recursively on $\{\bv_{i_0}, \bv_{i_1}, \dots, \bv_{i_r} \}$ where $\bv_{i_0} \in \bA_0$ and $\bv_{i_r} = \bv_j$ and get
\begin{align*}
    \bH_j \bwpa{j} &= \sum\limits_{s=i_1}^{i_{r-1}} \bK_{s}(j,s) \bK_{s}^{-1}(s,s) \be_s\\
    \mbox{E}[ \bw_j \given \bwpa{j} ] &= \sum\limits_{s=i_1}^{i_{r-1}} \mbox{E}_{\be_s}[\bw_j \given \be_s],
\end{align*}
where the expectations on the r.h.s. are taken with respect to the distributions of $\be_s$ which are Gaussian with mean zero and $\mbox{var}\{ \be_h\} = \bK_{h}(h,h)$ -- this is a compact expression of the conditionals governing the process as prescribed by $\calG$.
We can also write the above as $\mbox{E}(\bw_j \given \bwpa{j}) = \sum_{s=i_0}^{i_r} \bK_{s}(j,s) \bK_{s}^{-1}(s,s) (\bw_s - \mbox{E}[\bw_s \given \bwpa{s}])$; using $\mbox{E}[ \be_h \given \bw_h, \bw_{[h]}]=0$, for $h<k$ we find
\begin{align*}
    \mbox{cov}\{\be_h, \be_k \} &= \mbox{E}[ \mbox{cov}\{\be_h, \be_k \given \bw_h, \bw_{[h]}\}] + \mbox{cov}\{ \mbox{E}[\be_h \given \bw_h, \bw_{[h]}], \mbox{E}[ \be_k \given \bw_h, \bw_{[h]} ] \} \\
    &= \mbox{cov}\{ \mbox{E}[ \be_h \given \bw_h, \bw_{[h]}], \mbox{E}[ \be_k \given \bw_h, \bw_{[h]} ] \} = 0.
\end{align*}
The above results also imply $\Cov_{j,[j]}\Cov^{-1}_{[j]}\Cov_{[j],j} = \sum\limits_{s=i_0}^{i_r} \bK_{s}(j,s) \bK_{s}^{-1}(s,s) \bK_{s}(s,j)$ and suggest an additive representation via orthogonal basis functions:
\begin{equation}\label{equation:additive_repres}
    \bw_j = \sum\limits_{s=i_0}^{i_{r-1}} \bK_{s}(j,s) \bK_{s}^{-1}(s,s) \be_s + \be_j 
\end{equation} 

Finally, considering the same sequence of nodes, recursively introduce the covariance functions $\bF_0(r,s) = \Cov_{r,s}$ and for $j>1$, $\bF_j(r,s) = \bF_{j-1}(r,s) - \bF_{j\mbox{-}1}(r, j\mbox{-}1) \bF_{j\mbox{-}1}^{-1}(j\mbox{-}1, j\mbox{-}1) \bF_{j\mbox{-}1}(j\mbox{-}1, s)$. We get
\begin{align*}
    \bF_{j+1}(r, s) &= \bF_{j}(r,s)-\bF_j(r,j)\bF_j^{-1}(j,j) \bF_j(j, s)\\
  \text{\footnotesize using (\ref{equation:nested_inverse})}\quad  &= \bF_{j-1}(r,s) - \bF_{j-1}(r, [j\mbox{-1:}j])\bF_{j-1}^{-1}([j\mbox{-1:}j], [j\mbox{-1:}j]) \bF_{j-1}^{-1}([j\mbox{-1:}j], s) \\
    &= \Cov(r,s) - \Cov(r, [0\mbox{:}j]) \Cov^{-1}([0\mbox{:}j], [0\mbox{:}j]) \Cov([0\mbox{:}j], s)\\
    &= \Cov(r,s) - \Cov_{r,[j+1]} \Cov^{-1}_{[j+1]} \Cov_{[j+1],s}
\end{align*}
which can be iterated forward and results in an additional recursive way to compute covariances in \modelnames. Notice that while $\bK_j$ is formulated using the inverse of $J_{j}\times J_j$ matrix $\Cov_{[j]}$, the $\bF_j$'s require inversion of smaller $n_j \times n_j$ matrices $\bF_{j-1}(j\mbox{-}1, j\mbox{-}1)$.

\subsection{Properties of $\tilde{\Cov}$}
\subsubsection{\Large $\delta=1$} \label{section:covariance:limited}
Choosing depth $\delta=1$ results in each node having exactly $1$ parent. In this case the path $\calP_{k \to j} = \{\bv_{i_{1}}, \dots, \bv_{i_{r}} \}$ from $\bv_k$ to $\bv_j$, where $\bv_{i_{1}} = \bv_k$, $\bv_{i_{r}} = \bv_j$ and $\{ \bv_{i_h}\} = \pa{\bv_{i_{h+1}}}$, is unique, and there is thus no distinction between shortest and longest paths: $\calP_{k \to j}=\bar{\calP}_{k \to j}=\tilde{\calP}_{k \to j}$. Then denote $\dddot{\bH}_{k \to j} = \bH_{i_{r}} \cdot \bH_{i_{r-1}} \cdots \bH_{i_1}$. Let $\bv_z$ be the \textit{concestor} between $\bv_i$ and $\bv_j$ i.e. $\bv_z = \text{con}(\bv_i, \bv_j) = \arg \max_{\bv_k \in \bV} \{ k : \calP_{k\to i} \cap \calP_{k\to j} \neq \emptyset \}$ and the associated paths $\calP_{z \to i} = \{ \bv_{i_1}, \dots, \bv_{i_{r_i}} \}$ and $\calP_{z \to j} = \{ \bv_{j_1}, \dots, \bv_{j_{r_j}} \}$ where $\bv_{i_1} = \bv_{j_1} = \bv_z$, $\bv_{i_{r_i}} = \bv_i$ and $\bv_{j_{r_j}} = \bv_j$. Then we can write $\bw_i = \bw_{i_{r_i}} = \bH_{i_{r_i}} \bw_{i_{r_i-1}} + \bnu_{i_{r_i}}$ where $\bnu_{i_{r_i}} \sim N(\bzero, \bR_{i_{r_i}})$ and proceed expanding $\bw_{i_{r_i-1}}$ to get $\bw_{i_{r_i}} = \bH_{i_{r_i}} (\bH_{i_{r_i-1}} \bw_{i_{r_i-2}} + \bnu_{i_{r_i-1}}) + \bnu_{i_{r_i}} = \bH_{i_{r_i}} \bH_{i_{r_i-1}} \bw_{i_{r_i-2}} + (\bH_{i_{r_i}}\bnu_{i_{r_i-1}} + \bnu_{i_{r_i}})$; continuing downwardly along the tree we eventually find $\bw_i = \bH_{i_{r_i}} \cdots \bH_{i_1} \bw_{i_1} + \tilde{\bv}_i = \dddot{\bH}_{z \to i} \bw_z + \tilde{\bv}_i$ where $\tilde{\bv}_i$ is independent of $\bw_z$. After proceeding analogously with $\bw_j$, take $\bl_i, \bl_j $ such that $\eta(\bl_i)=\bv_i$ and $\eta(\bl_j) = \bv_j$. Then
\begin{equation}\label{equation:covariance_limited}
\text{Cov}_{\tp}(w(\bl_i), w(\bl_j)) = \dddot{\bH}_{z \to i}(\bl_i) \Cov_{z} \dddot{\bH}_{z\to j}(\bl_j)^{\top},
\end{equation}
where $\dddot{\bH}_{z\to i}(\bl_i) = \Cov(\bl_i, S_i) \Cov^{-1}_{i} \dddot{\bH}_{z \to {[i]}}$ and similarly for $\dddot{\bH}_{z\to j}(\bl_j)$.

\subsubsection{\Large $1<\delta<M$} \label{section:covariance:mixed}
Take two nodes $\bv_i, \bv_j \in \bV$. If $\pa{\bv_i} \cap \pa{\bv_j} \neq \emptyset$ then we apply the same logic as in \ref{section:covariance:full} using $\bv_z = \text{con}(\bv_i, \bv_j)$ as root. If $\pa{\bv_i} \cap \pa{\bv_j} = \emptyset$ and both nodes are at levels below $M_{\delta}$ then we use \ref{section:covariance:limited}. The remaining scenario is thus one in which $\bv_i \in \bA_r$, $r>M_{\delta}$ and $\pa{\bv_i} \cap \pa{\bv_j} = \emptyset$. We take $\bv_j \in \bA_s$, $s<M_{\delta}$ for simplicity in exposition and without loss of generality. By (\ref{equation:additive_repres})
\begin{equation}
\begin{aligned}
\bw_i &= \sum\limits_{s=i_{M_{\delta}}}^{i_{r-1}} \bK_{s}(i,s) \bK_{s}^{-1}(s,s) \be_s + \be_i,\\
&= \sum\limits_{s=i_{M_{\delta}+1}}^{i_{r-1}} \bK_{s}(i,s) \bK_{s}^{-1}(s,s) \be_s + 
\Cov_{ix} \Cov^{-1}_x \bw_x + \be_i,
\end{aligned}
\end{equation}
where $\bv_x \in \bA_{M_{\delta}}$ is the parent node of $\bv_i$ at level $M_{\delta}$. The final result of (\ref{equation:cov_limited}) is then achieved by noting that the relevant subgraph linking $\bv_x$ and $\bv_j$ has depth $\delta_x=1$ and thus $\text{Cov}(\bw_x, \bw_j)$ can be found via \ref{section:covariance:limited}, then $\text{Cov}(\bw_i, \bw_j) = \Cov_{ix} \Cov^{-1}_x \text{Cov}(\bw_x, \bw_j) = \bF_{i} \text{Cov}(\bw_x, \bw_j)$. Notice that $\bF_i$ directly uses the directed edge $\bv_x \to \bv_i$ in $\calG$; for this reason the path between $\bw_i$ and  $\bw_z=\text{con}(\bw_x, \bw_j)$ is the actually the shortest path and we have $\bv_z \to \cdots \to \bv_x \longrightarrow \bv_i$.

\subsubsection{\Large $\delta=M$}\label{section:covariance:full}
Take $\bv_i, \bv_j \in \bV$ and the full paths from the root $\tilde{\calP}_{0 \to i} = \{ i_0, \dots, i_{r_i} \}$ and $\tilde{\calP}_{0 \to j} = \{j_0, \dots, j_{r_j} \}$, respectively. Then using (\ref{equation:additive_repres}) we have 
\begin{equation} \label{equation:covariance_full_ex}
    \begin{aligned}
    \bw_i &= \sum\limits_{s \in \tilde{\calP}_{0\to i}} \bK_{s}(i,s) \bK_{s}^{-1}(s,s) \be_s + \be_i\\
        &= \sum\limits_{s \in \tilde{\calP}_{0\to i} \cap \tilde{\calP}_{0\to j}} \bK_{s}(i,s) \bK_{s}^{-1}(s,s) \be_s + \sum\limits_{s \in \tilde{\calP}_{0\to i} \setminus \tilde{\calP}_{0\to j}} \bK_{s}(i,s) \bK_{s}^{-1}(s,s) \be_s + \be_i\\
        &= \sum\limits_{s \in \tilde{\calP}_{0\to i} \cap \tilde{\calP}_{0\to j}} \bK_{s}(i,s) \bK_{s}^{-1}(s,s) \be_s + \tilde{\be}_i\\
    \bw_j &= \sum\limits_{s \in \tilde{\calP}_{0\to j}} \bK_{s}(j,s) \bK_{s}^{-1}(s,s) \be_s + \be_j \\
        &= \sum\limits_{s \in \tilde{\calP}_{0\to i} \cap \tilde{\calP}_{0\to j}} \bK_{s}(j,s) \bK_{s}^{-1}(s,s) \be_s + \sum\limits_{s \in \tilde{\calP}_{0\to j} \setminus \tilde{\calP}_{0\to i}} \bK_{s}(j,s) \bK_{s}^{-1}(s,s) \be_s + \be_j \\
        &= \sum\limits_{s \in \tilde{\calP}_{0\to i} \cap \tilde{\calP}_{0\to j}} \bK_{s}(j,s) \bK_{s}^{-1}(s,s) \be_s + \tilde{\be}_j,
    \end{aligned}
\end{equation}
where $\text{Cov}(\tilde{\be}_i, \tilde{\be}_j) = 0$. Then since $\be_s$ are independent and $\be_s \sim N(\bzero, \bK_s(s,s))$ we find
\begin{equation} \label{equation:covariance_full}
\begin{aligned}
\text{Cov}_{\tp}(\bw_i, \bw_j) &= \text{Cov}\left(\sum\limits_{s \in \tilde{\calP}_{0\to i} \cap \tilde{\calP}_{0\to j}} \bK_{s}(i,s) \bK_{s}^{-1}(s,s) \be_s + \be_i,\right.\\
&\left.\qquad \qquad \qquad \sum\limits_{s \in \tilde{\calP}_{0\to i} \cap \tilde{\calP}_{0\to j}} \bK_{s}(j,s) \bK_{s}^{-1}(s,s) \be_s+ \be_j \right)\\
&= \sum\limits_{s \in \tilde{\calP}_{0\to i} \cap \tilde{\calP}_{0\to j}} \bK_{s}(i,s) \bK_{s}^{-1}(s,s)\bK_{s}(s,j) + \bolds{1}_{i=j}\{ \bK_i(i,i) \}.
\end{aligned}
\end{equation}
We conclude by noting that $\delta=M$ implies $\tilde{\calP}_{0\to i} \cap \tilde{\calP}_{0\to j} = \pa{\bv_i} \cap \pa{\bv_j}$; considering two locations $\bl_i, \bl_j \in \calD^*$ such that $\eta(\bl_i) = \bv_i$ and $\eta(\bl_j) = \bv_{j}$ we obtain 
\begin{equation} \label{equation:covariance_full_location}
\begin{aligned}
\text{Cov}_{\tp}(w(\bl_i), w(\bl_j)) &= \sum\limits_{s : \{ \bv_s \in \pa{\bv_i} \cap \pa{\bv_j}\}} \bK_{s}(\bl_i,s) \bK_{s}^{-1}(s,s)\bK_{s}(s,\bl_j) + \bolds{1}_{\bl_i=\bl_j}\{ \bK_i(\bl_i,\bl_j) \}.
\end{aligned}
\end{equation}

\subsection{Computational cost}\label{appendix:computing_cost}
We make some assumptions here to simplify the calculation of overall cost: first, we assume that reference locations are all observed $\calS \subset \calT$, and consequently $\calU = \calT \setminus \calS$. Second, we assume that all reference subsets have the same size i.e. $|S_i| = N_s$ for all $i$. 
Third, we assume all nodes have the same number of children at the next level in $\calG$, i.e. if $\bv_i \in \bA_r$ with $r<M-1$, then $|\ch{\bv_i} \cap \bA_{r+1}| = C$, whereas if $r=M-1$ then $|\ch{\bv_i}|=N_u$.  Fourth, we assume that all non-reference subsets are singletons i.e. if $\bv_i \in \bB$ then $|U_i|=1$. The latter two assumptions imply (\ref{equation:u_locations_independent}). We also fix $C N_s = N_u$. As a result, the number of nodes at level $r=0, \dots, M-1$ is $C^r$, therefore $|\bA| + |\bB| = \sum_{r=0}^{M-1} C^r + N_u C^{M-1} = \frac{C^{M} - 1}{C-1} + N_s C^M$. Then the sample size is $n = |\calT| =  |\calS| + |\calU| = N_s \frac{C^{M+1}-1}{C-1}$ hence $M \approx \log_C (n/N_s)$.
Starting with $\delta=M$, the parent set sizes $J_i$ for a node $\bv_i \in \bA_r$ grow with $r$ as $J_i = r N_s$ and if $\bv_i \in \bB$ then $J_i = M N_s$. 
The cost of computing $p(\bw \given \btheta)$ is driven by the calculation of $\bH_j$, which is $O(r^2 N_s^3)$ for reference nodes at level $r$, for a total of $O(N_s^3 \sum_{r=0}^{M-1} C^r r^2)$. Since for common choices of $C$ and $M$ we have $\sum_{r=0}^{M-1} C^r r^2 N_s^3 \leq \sum_{r=0}^{M-1} C^{2r} N_s^3 = \frac{C^{2M}-1}{C^2-1} N_s^3 \approx C^M N_s^3 \approx \frac{n}{N_s} N_s^3 = n N_s^2$ then the cost for reference sets is $O(n N_s^2)$. Analogously for non reference nodes we get $O(C^{M} M^2 N_s^3)$ which leads to a cost of $O(n N_s^2)$. 
The cost of sampling $\bw$ is mainly driven by the computation of the Cholesky factor of a $N_s \times N_s$ matrix at each of $\frac{C^M-1}{C-1}$ reference nodes, which amounts to $O(n N_s^2)$. For the $N_s C^{M}$ non-reference nodes the main cost is in computing $\bH_i \bw_{[i]}$ which is $M^2 N_s^2$ for overall cost $O(C^{M} M^2 N_s^3)$ which again is $O(n N_s^2)$. Obtaining $\bF_i^{(c)}$ at the root of $\calG$ is associated to a cost $O(N_s^2 \frac{C^M - C}{C-1})$ which is $O(n N_s)$ but constitutes a bottleneck if such operation is performed simultaneously to sampling; however this bottleneck is eliminated in Algorithm \ref{algorithm:gibbs}.

If $\delta=1$ then the parent set sizes $J_i$ for all nodes $\bv_i \in \bV$ are constant $J_i = N_s$; since the nodes at levels $0$ to $M-1$ have $C$ children, the asymptotic cost of computing $p(\bw \given \btheta)$ is $O(N_s^3 \sum_{r=0}^{M-1} C^r) = O(N_s^3 \frac{C^{M}-1}{C-1}) = O(n N_s^2)$. However there are savings of approximately a factor of $M$ associated to $\delta=1$ in fixed samples since $\sum_{r=1}^{M-1} C^r r^2 > \sum_{r=1}^{M-1} C^r r > \frac{M C^M - 1}{C-1} - \frac{C^{M+1}}{(C-1)^2} > \frac{M C^M - 1}{C-1} > M \sum_{r=0}^{M-1} C^r$. Fixing $C$ and $M$ one can thus choose larger $N_s$ and smaller $\delta$, or vice-versa.

The storage requirements are driven by the covariance at parent locations $\Cov_{[j]}$ for nodes $\bv_j$ with $\pa{\bv_j}\neq \emptyset$ i.e. all reference nodes at level $r = 1, \dots, M-1$ and non-reference nodes. Taking $\delta=M$, suppose $\bv_i$ is the last parent of $\bv_j$, meaning $\bv_i \cup \pa{\bv_i} = \pa{\bv_j}$. Then $\Cov_{[j]} = \Cov(\{S_i, S_{[i]}\}, \{S_i, S_{[i]}\})$. If $\bv_i \in \bA_{r}$ then these matrices are of size $(r+1) N_s \times (r+1) N_s$; each of these is thus $O(r^2 N_s^2)$ in terms of storage. Considering all such matrices brings the overall storage requirement to $O(\sum_{r=0}^{M-1} C^r r^2 N_s^2)$ which is $O(n N_s)$ using analogous arguments as above. For $\delta=1$ we apply similar calculations as above. The same number of $\bH_j$ and $\bR_j$ must be stored but these are smaller in size and therefore do not affect the overall storage requirements. The design matrix $\bZ$ is stored in blocks and never as a large (sparse) matrix implying a storage requirement of $O(nq)$.

\section{Implementation details}\label{appendix:implementation}
Building a \modelname\ DAG proceeds by first constructing a base-tree $\calG_1$ at depth $\delta=1$ and then adding edges to achieve the desired depth level. The base tree $\calG_1$ is built from the root by branching each node $\bv$ into $|\ch{\bv}| = c^d$ children where $d$ is the dimension of the spatial domain and $c$ is a small integer. The spatial domain $\calD$ is partitioned recursively; after setting $\calD$, each recursive step proceeds by partitioning each coordinate axis of $D_i \subset \calD$ into $c$ intervals. As a consequence $D_i = \cup_j D_{ij}$ and $D_{ij} \cap D_{ij'} = \emptyset$ if $j\neq j'$. This recursive partitioning scheme is used to partition the reference set $\calS$ which we consider as a subset of the observed locations. Suppose we wish to associate node $\bv$ to approximately $n_S$ locations where $n_S = k^d$ for some $k$. Start from the root i.e. $\bv \in \bA_0$. Then take $\calS_{0} = \calS$ and partition it via parallel partitioning of each coordinate axis into $k$ intervals. Collect 1 location from each subregion to build $S_0$. Then set $\eta(S_0) = \bv_0$ and $\calS_1 = \calS \setminus \calS_0$. Then, take $\{ D_{1j} \}_j$ such that $\cup_j D_{1j} = D_0 = \calD$. We find $S_{1j}$ via axis-parallel partitioning of $\calS_1 \cap D_{1j}$ into $k^d$ regions and selecting one location from each partition, as above, and setting $\calS_2 = \calS \setminus \{ \calS_0 \cup \calS_1 \}$. All other reference subsets are found by sequentially removing locations from the reference set, and proceeding analogously as above. This stepwise procedure is stopped when either the tree reaches a predetermined height $M$, or when there is an insufficient number of remaining locations to build reference subsets of size $n_S$. The remaining locations are assigned to the leaf nodes via $\eta_B$ as defined in Section \ref{section:construction_parts} in order to include at least one neighboring realization of the process from the same variable. 

One specific issue arises when multivariate data are imbalanced, i.e. one of the margins is observed at a much sparser grid, e.g. in Section \ref{section:applications:modisnoaa} \texttt{PRCP} is collected at a ratio of 1:10 locations compared to other variables. In these cases, if locations were chosen uniformly at random to build the reference subsets then the root nodes would be associated via $\eta$ to reference subsets which likely do not contain such sparsely observed variables. This scenario goes against the intuition of \ref{prop:prop1} suggesting that a na\"ive approach would result in poor performance at the sparsely-observed margins. To avoid such a scenario, we bias the sampling of locations to favor those at which the sparsely-observed variables are recorded. As a result, in Section \ref{section:applications:modisnoaa} near-root nodes are associated to reference subsets in which all variables are balanced; the imbalances of the data are reflected by imbalanced leaf nodes instead.

The source code for \modelnames\ is available at \repourl\ and can be installed as an \texttt{R} package. The \texttt{spamtree} package is written in C++ using the Armadillo library for linear algebra \citep{armadillo} interfaced to \texttt{R} via \texttt{RcppArmadillo} \citep{rcpparmadillo}. All matrix operations are performed efficiently by linkage to the LAPACK and BLAS libraries \citep{blackford2002blas, lapack99} as implemented in OpenBLAS 0.3.10 \citep{openblas} or the Intel Math Kernel Library. Multithreaded operations proceed via OpenMP \citep{dagum1998openmp}.

\subsection{Applications}
\subsubsection{Simulated datasets} \label{appendix:synth_comparison}
\begin{figure}
    \centering
    \includegraphics[width=.95\textwidth]{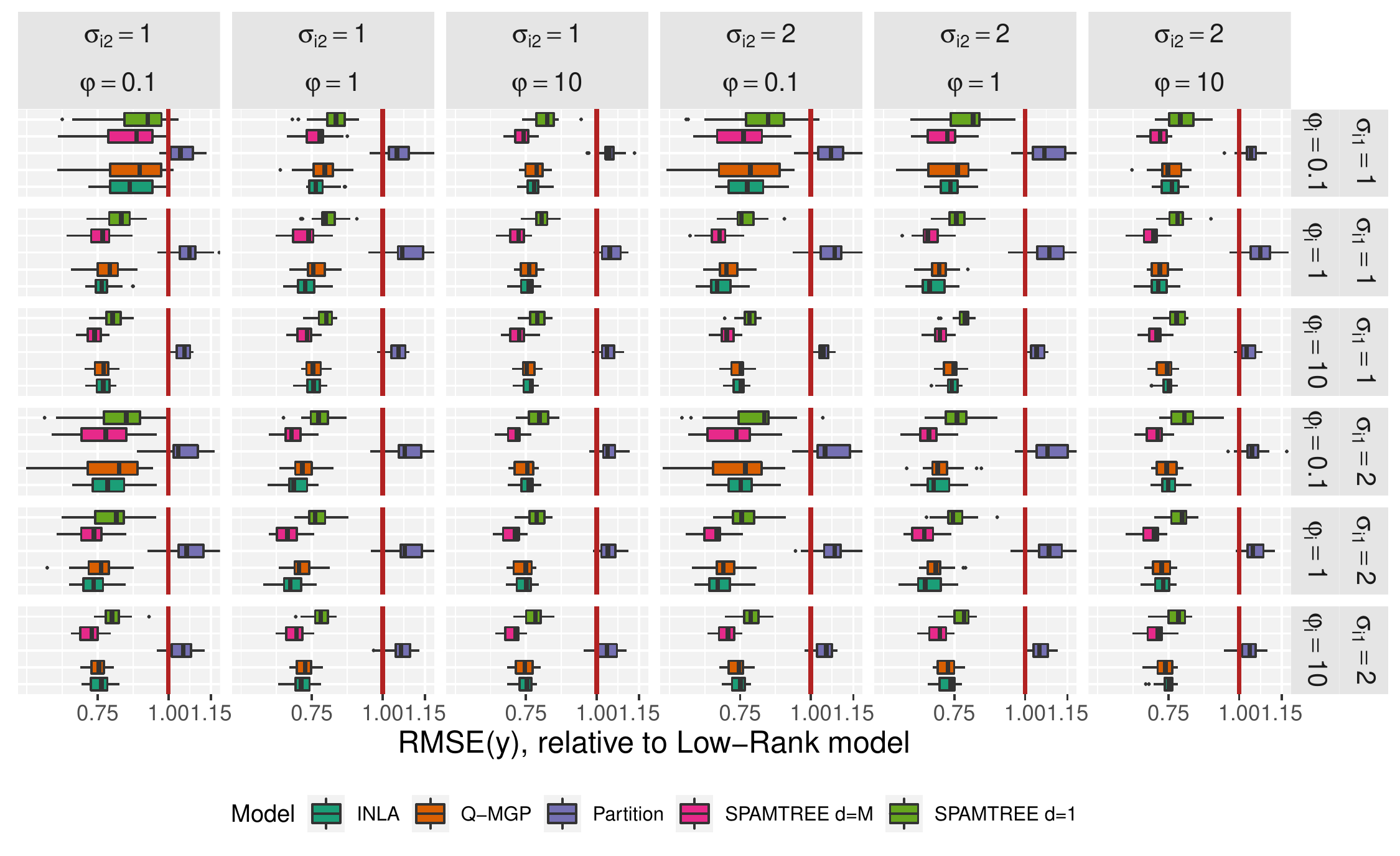}
    \caption{RMSE in out-of-sample predictions for spatial models and each setting of $\sigma_{i1}, \sigma_{i2}, \phi_i, \phi$, with $i=1,2$, relative to the RMSE of the low rank GP method.}
    \label{figure:synthetic_boxplots_mae}
\end{figure}

\begin{figure}
    \centering
    \includegraphics[width=.95\textwidth]{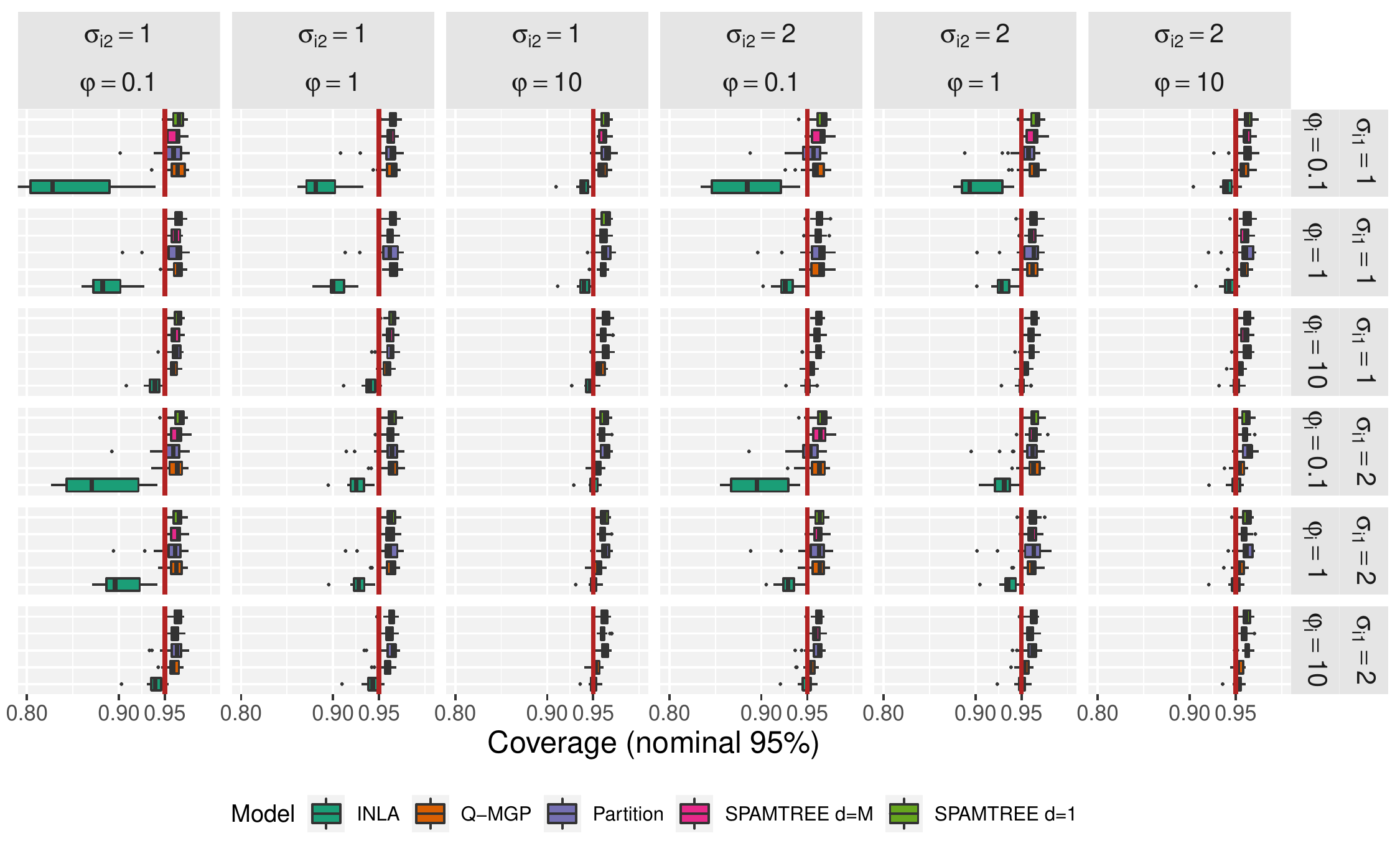}
    \caption{Coverage in out-of-sample predictions for spatial models and each setting $\sigma_{i1}, \sigma_{i2}, \phi_i, \phi$.}
    \label{figure:synthetic_boxplots_covg}
\end{figure}

Multivariate \modelnames\ with full depth are implemented by targeting reference subsets of size $n_S = 25$ and tress with $c=4$ additional children for each branch. The tree is built starting from a $2 \times 2$ partition of the domain, hence there are 4 root nodes with no parents in the DAG.
The cherry-pickying function $\eta$ is set as in Section \ref{section:construction_parts}; with these settings the tree height is $M=3$. For \modelnames\ with depth $\delta=1$ we build the tree with reference subsets of size $n_S = 80$ and $c=4$. 
Multivariate MGPs are implemented via axis-parallel partitioning using 57 intervals along each axis. Multivariate INLAs based on the stochastic partial differential equation representation of GMRFs \citep{spde} were implemented following the examples in \cite{inlabook}, Chapter 3, setting the grid size to $15 \times 15$ to limit the compute time to 15 seconds when using 10 CPU threads. BART was implemented on each dataset via the $\texttt{wbart}$ function in the R package \texttt{BART}; the set of covariates for BART was built using the spatial coordinates in addition to a binary variable representing the output variable index (i.e. taking value 1 whenever $y_i$ is of the first outcome variable, 0 otherwise).

\subsubsection{MODIS-TERRA and GHCN}\label{appendix:modisnoaa}
The implemented \modelnames\ are built with 36 root nodes and $c=6$ additional children for each level of the tree, for up to $M=5$ levels of the tree and $\delta=5$ (i.e. full depth).
The non-reference observed locations are linked to leaves via cherry-pickying as in Section \ref{section:construction_parts}. The analysis was run on an AMD Epyc 7452-based virtual machine in the Microsoft Azure cloud; the \modelname\ R package was set to run on 20 CPU threads, on R version 4.0.3 linked to the Intel Math Kernel Library (MKL) version 2019.5-075.

\begin{figure}
    \centering
    \begin{tabular}{c}
    \resizebox{.90\columnwidth}{!}{%
\begin{tabular}{ccccc}
\multicolumn{5}{c}{} \\
\multicolumn{4}{c}{ \begin{tabular}{|r|ccc|}
  \hline
  $i$ & \large $\sigma_{i1}$ \normalsize & \large $\sigma_{i2}$ \normalsize & \large $\phi_i$ \normalsize \\ 
  \hline
  \footnotesize \texttt{LST\_Day\_CMG} & \renewcommand{\arraystretch}{0.3} 
  \begin{tabular}{@{}c@{}}$-0.8936$ \\ {\footnotesize $-0.9499, -0.8401$ } \end{tabular}\renewcommand{\arraystretch}{1} & \renewcommand{\arraystretch}{0.3} 
  \begin{tabular}{@{}c@{}}$8.3285$ \\ {\footnotesize $7.8071, 8.9614$ } \end{tabular}\renewcommand{\arraystretch}{1} & \renewcommand{\arraystretch}{0.3} 
  \begin{tabular}{@{}c@{}}$0.2174$ \\ {\footnotesize $0.1854, 0.2460$ } \end{tabular}\renewcommand{\arraystretch}{1} \\
   \footnotesize  \texttt{LST\_Night\_CMG} & \renewcommand{\arraystretch}{0.3} 
    \begin{tabular}{@{}c@{}}$-1.4104$ \\ {\footnotesize $-1.4794, -1.3530$ } \end{tabular}\renewcommand{\arraystretch}{1} & \renewcommand{\arraystretch}{0.3} 
    \begin{tabular}{@{}c@{}}$7.3927$ \\ {\footnotesize $7.0730, 7.7230$ } \end{tabular}\renewcommand{\arraystretch}{1} & \renewcommand{\arraystretch}{0.3} 
    \begin{tabular}{@{}c@{}}$0.0968$ \\ {\footnotesize $0.0883, 0.1054$ } \end{tabular}\renewcommand{\arraystretch}{1} \\
  \footnotesize \texttt{Clear\_sky\_days} & \renewcommand{\arraystretch}{0.3} 
  \begin{tabular}{@{}c@{}}$0.9189$ \\ {\footnotesize $0.8695, 0.9708$ } \end{tabular}\renewcommand{\arraystretch}{1} & \renewcommand{\arraystretch}{0.3} 
  \begin{tabular}{@{}c@{}}$3.3133$ \\ {\footnotesize $3.3033, 3.4523$ } \end{tabular}\renewcommand{\arraystretch}{1} & \renewcommand{\arraystretch}{0.3} 
  \begin{tabular}{@{}c@{}}$0.5790$ \\ {\footnotesize $0.5303, 0.6185$ } \end{tabular}\renewcommand{\arraystretch}{1} \\
   \footnotesize  \texttt{Clear\_sky\_nights} & \renewcommand{\arraystretch}{0.3} 
    \begin{tabular}{@{}c@{}}$3.8138$ \\ {\footnotesize $3.7138, 3.9306$ } \end{tabular}\renewcommand{\arraystretch}{1} & \renewcommand{\arraystretch}{0.3} 
    \begin{tabular}{@{}c@{}}$0.9603$ \\ {\footnotesize $0.9194, 1.0114$ } \end{tabular}\renewcommand{\arraystretch}{1} & \renewcommand{\arraystretch}{0.3} 
    \begin{tabular}{@{}c@{}}$6.2129$ \\ {\footnotesize $5.7944, 6.6519$ } \end{tabular}\renewcommand{\arraystretch}{1} \\
    \texttt{PRCP} & \renewcommand{\arraystretch}{0.3} 
    \begin{tabular}{@{}c@{}}$-0.3009$ \\ {\footnotesize $-0.3348, -0.2702$ } \end{tabular}\renewcommand{\arraystretch}{1} & \renewcommand{\arraystretch}{0.3} 
    \begin{tabular}{@{}c@{}}$0.6897$ \\ {\footnotesize $0.6466, 0.7200$ } \end{tabular}\renewcommand{\arraystretch}{1} & \renewcommand{\arraystretch}{0.3} 
    \begin{tabular}{@{}c@{}}$0.1832$ \\ {\footnotesize $0.1655, 0.2051$ } \end{tabular}\renewcommand{\arraystretch}{1} \\
    \hline
\end{tabular} } & 
\begin{tabular}{|c|}
  \hline
  \large $\alpha$ \normalsize \\
  \renewcommand{\arraystretch}{0.3} \begin{tabular}{@{}c@{}}$0.1012$ \\ {\footnotesize $0.0696, 0.1248$ } \end{tabular}\renewcommand{\arraystretch}{1} \\
  \hline
  \large $\beta$ \normalsize \\
  \renewcommand{\arraystretch}{0.3} \begin{tabular}{@{}c@{}}$0.1654$ \\ {\footnotesize $0.1258, 0.2203$ } \end{tabular}\renewcommand{\arraystretch}{1} \\
  \hline
  \large $\phi$ \normalsize \\
  \renewcommand{\arraystretch}{0.3} \begin{tabular}{@{}c@{}}$0.5715$ \\ {\footnotesize $0.5326, 0.6079 $ } \end{tabular}\renewcommand{\arraystretch}{1}\\
  \hline
\end{tabular} \\
\multicolumn{5}{c}{ }\\
\multicolumn{5}{c}{\begin{tabular}{|r|cccc|}
    \hline
    $\delta_{ij}$ & \rotatebox{-0}{\footnotesize \texttt{LST\_Day\_CMG}} & \rotatebox{-0}{\footnotesize \texttt{LST\_Night\_CMG}} & \rotatebox{-0}{\footnotesize \texttt{Clear\_sky\_days}} & \rotatebox{-0}{\footnotesize \texttt{Clear\_sky\_nights}} \\
        \hline
        \multicolumn{1}{|r|}{\footnotesize\texttt{LST\_Night\_CMG}} & \renewcommand{\arraystretch}{0.3} \begin{tabular}{@{}c@{}}$0.1279$ \\ {\footnotesize $0.0608, 0.2328$ } \end{tabular}\renewcommand{\arraystretch}{1}  & & & \multicolumn{1}{c|}{ } \\ 
        \multicolumn{1}{|r|}{\footnotesize\texttt{Clear\_sky\_days}} & \renewcommand{\arraystretch}{0.3} \begin{tabular}{@{}c@{}}$1.7295$ \\ {\footnotesize $1.6639, 1.7962$ }  \end{tabular}\renewcommand{\arraystretch}{1} & \renewcommand{\arraystretch}{0.3} 
        \begin{tabular}{@{}c@{}}$1.5371$ \\ {\footnotesize $1.3765, 1.7059$ }  \end{tabular}\renewcommand{\arraystretch}{1}  & & \multicolumn{1}{c|}{ }\\
        \multicolumn{1}{|r|}{\footnotesize\texttt{Clear\_sky\_nights}} & \renewcommand{\arraystretch}{0.3} \begin{tabular}{@{}c@{}}$0.0307$ \\ {\footnotesize $0.0221, 0.0395$ }  \end{tabular}\renewcommand{\arraystretch}{1} & \renewcommand{\arraystretch}{0.3} 
        \begin{tabular}{@{}c@{}}$1.1156$ \\ {\footnotesize $0.8964, 1.3194$ }  \end{tabular}\renewcommand{\arraystretch}{1} & \renewcommand{\arraystretch}{0.3} 
        \begin{tabular}{@{}c@{}}$1.5035$ \\ {\footnotesize $1.2670, 1.7380$ }  \end{tabular}\renewcommand{\arraystretch}{1} & \multicolumn{1}{c|}{ } \\
        \multicolumn{1}{|r|}{\texttt{PRCP}} & \renewcommand{\arraystretch}{0.3} 
        \begin{tabular}{@{}c@{}}$0.2436$ \\ {\footnotesize $0.2039, 0.2878$ }  \end{tabular}\renewcommand{\arraystretch}{1} & \renewcommand{\arraystretch}{0.3} 
        \begin{tabular}{@{}c@{}}$1.3151$ \\ {\footnotesize $0.9149, 1.7000$ }  \end{tabular}\renewcommand{\arraystretch}{1} & \renewcommand{\arraystretch}{0.3} 
        \begin{tabular}{@{}c@{}}$0.0572$ \\ {\footnotesize $0.0490, 0.0643$ }  \end{tabular}\renewcommand{\arraystretch}{1} & \multicolumn{1}{c|}{ \renewcommand{\arraystretch}{0.3} 
        \begin{tabular}{@{}c@{}}$0.7677$ \\ {\footnotesize $0.4010, 1.1468$ }  \end{tabular}\renewcommand{\arraystretch}{1}} \\
        \hline
    \end{tabular}}
\end{tabular}
} \end{tabular}
    \caption{Posterior means and 95\% credible intervals for components of $\btheta$ for \modelnames.}
    \label{fig:covariance_table}
\end{figure}

Figure \ref{fig:meshgp_res} reports predictive performance of a tessellated MGP \citep{meshedgp} implemented on the same data; it can be compared to Figure \ref{fig:results_plot} in the main article. The MGP model was implemented via the development package at \url{github.com/mkln/meshgp} targeting a block size with 4 spatial locations, resulting in an effective average block dimension of 20. Caching was unavailable due to the irregularly spaced \texttt{PRCP} values. Fewer MCMC iterations were run compared to \modelnames\ to limit total runtime to less than 16h. 

\begin{figure}
    \centering
    \resizebox{.90\columnwidth}{!}{%
    \begin{tabular}{|r|ccccc|}
  \hline
  Measure & 
  \footnotesize \texttt{Clear\_sky\_days} & 
  \footnotesize\texttt{Clear\_sky\_nights} & 
  \footnotesize\texttt{LST\_Day\_CMG} &
  \footnotesize \texttt{LST\_Night\_CMG} & 
  \texttt{PRCP} \\ 
  \hline
    95\% Coverage & 0.8662 & 0.9427 & 1.0000 & 0.9991 & 1.0000 \\ 
    MAE & 1.5301 & 1.3935 & 1.3507 & 1.1653 & 0.4902 \\ 
    RMSE & 1.9276 & 1.7664 & 1.6991 & 1.4024 & 0.6315 \\ 
   \hline
\end{tabular}}
\\
\resizebox{.915\columnwidth}{!}{
\begin{tabular}{|c|c|c|c|}
\hline
    $n=$1,014,017 &  Total iterations: 20,000 & Total time: 15.64h &
    Average time/iteration: 2.8s \\
    \hline
\end{tabular}
}
\caption{Prediction results over the $3 \times 3$ degree area shown in Figure \ref{fig:usamap} for a tessellated MGP.}
\label{fig:meshgp_res}
\end{figure}

\bibliographystyle{hapalike} 
\bibliography{biblio}

\end{document}